\documentclass[aps, prl,superscriptaddress,floatfix,showpacs,notitlepage,twocolumn,longbibliography]{revtex4-2}
\usepackage{color}
\usepackage{mathtools}
\usepackage{graphicx}
\usepackage{hyperref}
\usepackage{bbold}
\usepackage{braket}
\usepackage{microtype}
\usepackage{siunitx}
\usepackage{soul}
\usepackage{verbatim}
\usepackage{dsfont} 
\usepackage{bm}
\usepackage{tabularx}
\usepackage{hhline}
\usepackage{booktabs}
\usepackage{makecell}
\usepackage{multirow}

\hypersetup{
  linkcolor = black,
  citecolor  = blue,
  urlcolor = blue,
  colorlinks = true,
}

\def\be{\begin{equation}}
\def\ee{\end{equation}}
\def\bea{\begin{eqnarray}}
\def\eea{\end{eqnarray}}
\def\Tr{{\rm Tr}}

\begin{document}

\title{Algorithmic  Quantum Simulations of Quantum Thermodynamics}

\newcommand{\affilUSTCHeFei}{Hefei National Research Center for Physical Sciences at the Microscale and School of Physical Sciences, University of Science and Technology of China, Hefei 230026, China}
\newcommand{\affilUSTCShanghai}{Shanghai Research Center for Quantum Science and CAS Center for Excellence in Quantum Information and Quantum Physics, University of Science and Technology of China, Shanghai 201315, China}
\newcommand{\affilFudan}{State Key Laboratory of Surface Physics, Institute of Nanoelectronics and Quantum Computing, and Department of Physics, Fudan University, Shanghai 200433, China}
\newcommand{\affilQiZhi}{Shanghai Qi Zhi Institute, AI Tower, Xuhui District, Shanghai 200232, China}
\newcommand{\affilShanghaiQuantum}{Shanghai Research Center for Quantum Sciences, Shanghai 201315, China}
\newcommand{\affilInnsbruckTheory}{Institute for Theoretical Physics, University of Innsbruck, 6020 Innsbruck, Austria}
\newcommand{\affilInnsbruckIQOQI}{Institute for Quantum Optics and Quantum Information of the Austrian Academy of Sciences, 6020 Innsbruck, Austria}
\newcommand{\affilHefeiNatLab}{Hefei National Laboratory, University of Science and Technology of China, Hefei 230088, China}
\newcommand{\affilJinan}{Jinan Institute of Quantum Technology and Hefei National Laboratory Jinan Branch, Jinan 250101, China}

\author{Yangsen Ye}
\altaffiliation{These authors contributed equally to this work.}
\affiliation{\affilUSTCHeFei}
\affiliation{\affilUSTCShanghai}

\author{Jue Nan}
\altaffiliation{These authors contributed equally to this work.}
\affiliation{\affilFudan}
\affiliation{\affilQiZhi}
\affiliation{\affilShanghaiQuantum}

\author{Dong Chen}
\altaffiliation{These authors contributed equally to this work.}
\affiliation{\affilFudan}
\affiliation{\affilQiZhi}
\affiliation{\affilShanghaiQuantum}

\author{Torsten V. Zache}
\affiliation{\affilInnsbruckTheory}
\affiliation{\affilInnsbruckIQOQI}

\author{Qingling Zhu}
\affiliation{\affilUSTCShanghai}
\affiliation{\affilHefeiNatLab}

\author{Yiming Zhang}
\affiliation{\affilUSTCHeFei}
\affiliation{\affilUSTCShanghai}

\author{Yuan Li}
\affiliation{\affilUSTCHeFei}
\affiliation{\affilUSTCShanghai}

\author{Xiawei Chen}
\affiliation{\affilUSTCShanghai}

\author{Chong Ying}
\affiliation{\affilUSTCShanghai}
\affiliation{\affilHefeiNatLab}

\author{Chen Zha}
\affiliation{\affilUSTCShanghai}
\affiliation{\affilHefeiNatLab}

\author{Sirui Cao}
\affiliation{\affilUSTCHeFei}
\affiliation{\affilUSTCShanghai}

\author{Shaowei Li}
\affiliation{\affilUSTCShanghai}
\affiliation{\affilHefeiNatLab}

\author{Shaojun Guo}
\affiliation{\affilUSTCHeFei}
\affiliation{\affilUSTCShanghai}

\author{Haoran Qian}
\affiliation{\affilUSTCHeFei}
\affiliation{\affilUSTCShanghai}

\author{Hao Rong}
\affiliation{\affilUSTCHeFei}
\affiliation{\affilUSTCShanghai}

\author{Yulin Wu}
\affiliation{\affilUSTCHeFei}
\affiliation{\affilUSTCShanghai}

\author{Kai Yan}
\affiliation{\affilUSTCShanghai}
\affiliation{\affilHefeiNatLab}

\author{Feifan Su}
\affiliation{\affilUSTCShanghai}
\affiliation{\affilHefeiNatLab}

\author{Hui Deng}
\affiliation{\affilUSTCHeFei}
\affiliation{\affilUSTCShanghai}
\affiliation{\affilHefeiNatLab}

\author{Yu Xu}
\affiliation{\affilUSTCShanghai}
\affiliation{\affilHefeiNatLab}

\author{Jin Lin}
\affiliation{\affilUSTCShanghai}
\affiliation{\affilHefeiNatLab}

\author{Ming Gong}
\affiliation{\affilUSTCHeFei}
\affiliation{\affilUSTCShanghai}
\affiliation{\affilHefeiNatLab}

\author{Fusheng Chen}
\affiliation{\affilUSTCShanghai}
\affiliation{\affilHefeiNatLab}

\author{Gang Wu}
\affiliation{\affilUSTCHeFei}
\affiliation{\affilHefeiNatLab}

\author{Yong-Heng Huo}
\affiliation{\affilUSTCHeFei}
\affiliation{\affilUSTCShanghai}
\affiliation{\affilHefeiNatLab}

\author{Chao-Yang Lu}
\affiliation{\affilUSTCHeFei}
\affiliation{\affilUSTCShanghai}
\affiliation{\affilHefeiNatLab}

\author{Cheng-Zhi Peng}
\affiliation{\affilUSTCHeFei}
\affiliation{\affilUSTCShanghai}
\affiliation{\affilHefeiNatLab}

\author{Xiaobo Zhu}
\affiliation{\affilUSTCHeFei}
\affiliation{\affilUSTCShanghai}
\affiliation{\affilHefeiNatLab}
\affiliation{\affilJinan}

\author{Xiaopeng Li}
\email{xiaopeng_li@fudan.edu.cn}
\affiliation{\affilFudan}
\affiliation{\affilQiZhi}
\affiliation{\affilShanghaiQuantum}
\affiliation{\affilHefeiNatLab}

\author{Jian-Wei Pan}
\email{pan@ustc.edu.cn}
\affiliation{\affilUSTCHeFei}
\affiliation{\affilUSTCShanghai}
\affiliation{\affilHefeiNatLab}

\begin{abstract}
Characterizing quantum phases-of-matter at finite-temperature is essential for understanding complex materials and large-scale thermodynamic phenomena.  Here, we develop algorithmic protocols for simulating quantum thermodynamics on quantum hardware through quantum kernel function expansion (QKFE), producing the free energy as an analytic function of temperature with uniform convergence. 
These protocols are demonstrated by simulating transverse field Ising and XY models with superconducting qubits. 
In both analogue and digital implementations of the QKFE algorithms, we exhibit quantitative agreement of our quantum simulation experiments with the exact results. 
Our approach provides a general framework for computing thermodynamic potentials on programmable quantum devices, granting access to key thermodynamic properties such as entropy, heat capacity and criticality, with far-reaching implications for material design and drug development.
\end{abstract}

\maketitle
\setlength{\parskip}{0.5\baselineskip}

\paragraph*{Introduction.---}
Simulating quantum many-body systems at finite temperature ({\bf T}) is pivotal across scientific domains, from high-energy and condensed matter physics to computer-assisted drug design.
Given that absolute zero temperature is unattainable for macroscopic systems, most of the physical processes of interest occur at finite temperatures, underscoring the significance of studying quantum thermodynamics.

Classical computing, while successful in specific regimes (e.g. Monte Carlo~\cite{1986_Ceperley_Science} and tensor networks~\cite{2021_Verstraete_RMP}), faces fudamental bottlenecks arsing from the notorious sign problem and strong quantum entanglement.
Kernel polynomial methods (KPM) express the thermodynamic properties as a series expansion in a complete function basis, reducing the challenge of diagonalizing a large Hamiltonian matrix to iterative matrix multiplication when treating finite energy density problems~\cite{RevModPhys.78.275}. However, computing the expansion coefficients remains classically intractable due to the exponential growth of Hilbert space.

\begin{figure}[htp]
    \centering
    \includegraphics[width=0.48\textwidth]{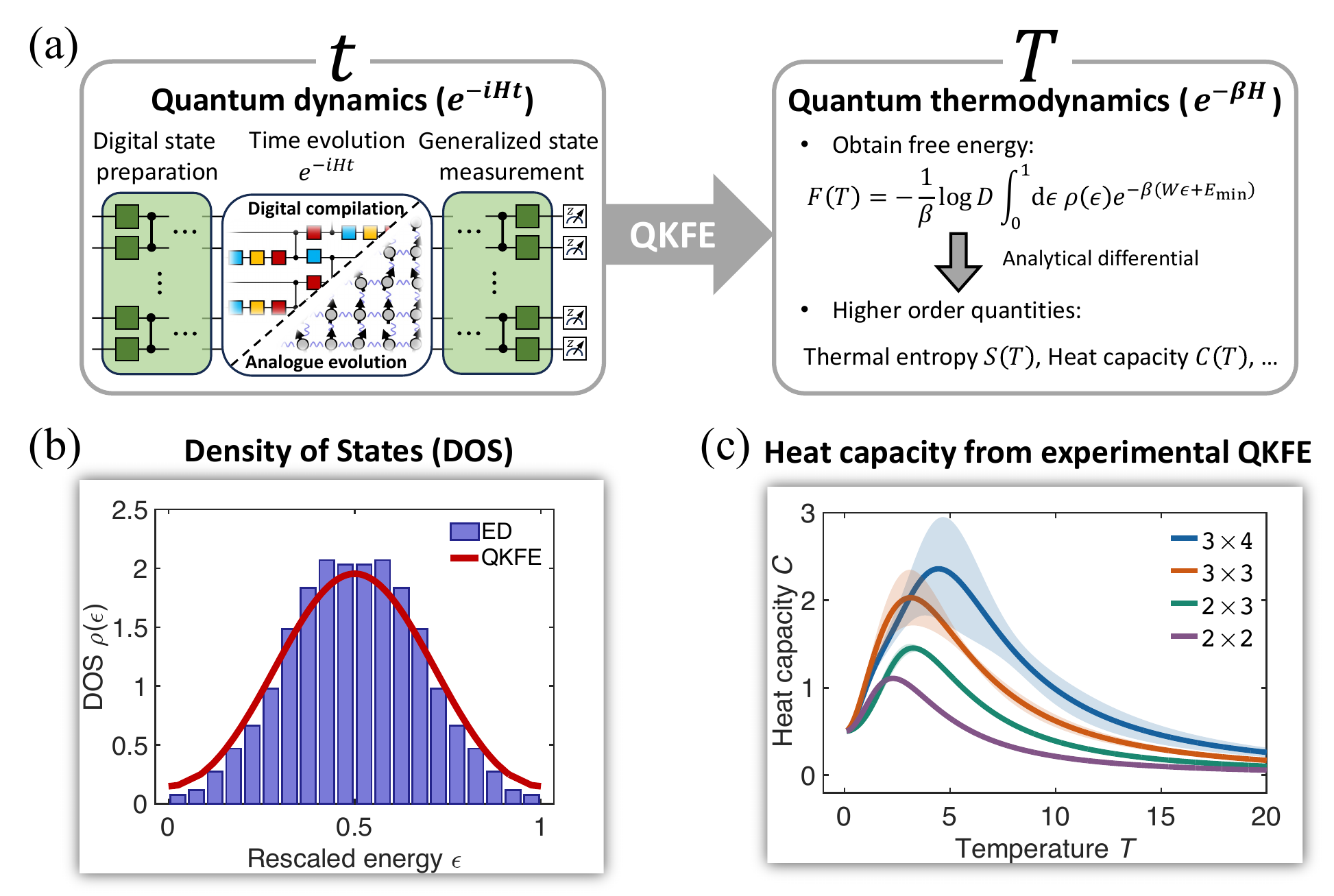}
    \caption{
    Schematic of the Quantum Kernel Function Expansion (QKFE) algorithm for simulating quantum thermodynamics.
    (a) The QKFE converts real-time (t) quantum dynamics  into finite-temperature (T) quantum thermodynamics by computing the free energy $F(T)$. Higher order quantities such as the entropy $S(T)$ and the heat capacity $C(T)$ are subsequently obtained via analytic differentiation.
    (b) DOS for the 1D 10-qubit TFIM: comparing the QKFE approximation (red line) with exact diagonalization (ED, histogram).
    (c) The extensive scaling of the heat capacity in the 2D TFIM obtained by implementing the QKFE algorithm on the quantum processor.
    The shaded areas represent the error bars corresponding to one standard deviation. Notably, for the $2\times 2$ and $2 \times 3$ cases, the error bars are smaller than the linewidth at this scale, a result of extensive averaging over samples and quantum shots.
    }
    \label{fig:Protocal}
\end{figure}

Quantum computing offers a transformative alternative by exploiting quantum superposition, enabling the direct manipulation of this complexity~\cite{2018_Preskill_Quantum,2022_Bharti_RMP,2022_Zoller_Nature, ZNE1experiment}.
Rapid developments in quantum hardware now allow for precise programming of system couplings and controllable evolution in real time ({\bf t}), fundamentally avoiding the sign problem~\cite{2013_Esslinger_Science,2015_Hulet_Nature,2017_Greiner_Nature,2024_Pan_Nature, 2001_Farhi_Science,2022_Lukin_Science,2024_Dwave_Nature,2014_Guzik_Nature,Patrick2021NatureReview,2024_Gambetta_Nature, 2020_Chan_NatPhys,2021_Minnich_PRXQ, CiracPRXQ2021,2024_Cirac_Quantum,RamilPRXQ2024}.
Leveraging these capabilities, the expansion moments in the KPM can be estimated on a programmable quantum device with polynomial measurement overhead~\cite{QKFE}.

In this work, we develop algorithmic protocols for computing equilibrium thermodynamic potentials with a QKFE method~\cite{QKFE}, 
{a KPM~\cite{RevModPhys.78.275} inspired quantum algorithm.}  
Building on the unique capability of programmable quantum devices to implement unitary evolution $e^{-iHt}$, generated by Hamiltonian $H$, the QKFE translates non-equilibrium measurement data at some time $t$ into equilibrium thermodynamics for a canonical Gibbs state $ e^{-H/k_BT}$, effectively establishing the mapping: $t \rightarrow T$ (Fig.~\ref{fig:Protocal}). 
We experimentally demonstrate QKFE in both digital~\cite{2024_Good_PRR} and analogue quantum simulations.
Our digital quantum simulations of transverse field Ising model (TFIM) accurately captures the quantum ferromagnetic-paramagnetic duality in one-dimension and critical behaviors in two-dimensions.
As a complementary example, we simulate the spin XY model with an analogue approach.
This underscores the versatility of our algorithmic quantum simulation approach to quantum thermodynamics, opening up novel opportunities for finite-temperature problems in many-body physics and quantum chemistry.

\paragraph*{Thermodynamics from algorithmic quantum simulation protocols.---}
The free energy is a fundamental thermodynamic quantity in describing quantum thermodynamics. 
Represented as  a function of temperature and system size, $T$ and $L$, the free energy function, $F(T, L)$, serves as a thermodynamic potential, yielding crucial thermodynamic information via differentiation~\cite{pathria2017statistical}.  
For a quantum system governed by Hamiltonian $H$, the QKFE algorithm estimates $F$ by leveraging the quantum dynamics $e^{-iHt}$, executed on a quantum processor. This is achieved by approximating the density of states (DOS) using a cosine expansion~\cite{QKFE}
\be
\rho(\epsilon) = f_0 +2\sum_{n>0} f_n\cos(n\pi\epsilon),
\ee
where $\epsilon \in [0,1]$ is the rescaled energy.
The expansion moments are given by the trace of the rescaled time-evolution operator as
\be
f_{n=1,2,\ldots} = {\rm Re} \left\{  \Tr [e^{-in \pi\tilde{H} } ] \right\} /D. 
\ee  
Here, $D$ is the Hilbert space dimension (which implies $f_0=1$), and $\tilde{H} = (H-E_\text{min}\mathbb{1})/W$ is the dimensionless Hamiltonian, with $W$ and $E_\text{min}$ estimates of the spectra width, and the ground state energy, respectively.
We emphasize these quantities are employed solely to rescale the spectrum into the interval $[0,1]$; consequently, precise determination is not required.
The free energy can then be given by 
\be\label{eq:freeEnergy}
\begin{aligned}
&F(T,L) = -k_\text{B} T \ln \left[ D \sum_{n>0} f_n \phi_n (\beta W) \right] + E_\text{min}, \\
&\phi_0(x) = \frac{1-e^{-x}}{x},\ \phi_n(x) = \frac{2[1-(-1)^n e^{-x}]}{x + n^2\pi^2/x},  
\end{aligned}
\ee
with $\beta$ the inverse temperature, $x=W/k_\text{B}T$.
Partition function is yielded via $Z=e^{-\beta F}$.
We thus obtain an approximation of $F$ by truncating the series in Eq.(\ref{eq:freeEnergy}) at a finite order $N$. To suppress the Gibbs oscillations arising from this finite cutoff, the expansion moments are corrected by multiplying a Jackson kernel~\cite{QKFE}.
It can be shown that the kernel function expansion has a uniform convergence with a bounded error $\omega_\kappa(1/N)$,
with $\omega_\kappa(\delta)$ defined as max$|\kappa(x)-\kappa(y)|_{|x-y|\leq\delta}$.

One straightforward way to implement the QKFE algorithm is to construct a controlled-unitary quantum circuit~\cite{QKFE}, 
$
{\cal U} = e^{-in\pi \ket{0}  \bra{0} \otimes \tilde{H}} 
$
,which acts on an ancilla qubit and $L$ physical qubits. 
Direct compilation of the controlled-unitary requires extensive qubit swap operations 
or non-local two-qubit gates, posing significant challenges for scaling to large systems. Estimating the density-of-states has been previously demonstrated in this way by exploiting the non-local interactions with trapped ions~\cite{2024_Good_PRR}, but it remains an outstanding challenge to construct a scalable quantum protocol and  precisely determine the free energy, 
which is the key for simulating quantum thermodynamics. 
To reach a scalable quantum algorithm on quantum processors with local connectivity, in the next section, we develop efficient compiling protocols that remove the requirement of non-local operations, assuming the Hamiltonian obeys certain symmetries.

\paragraph*{Efficient compilation via Hamiltonian symmetries.---}
Firstly, we consider an anti-commuting symmetry, 
$
\left\{ {\tau},  \, H \right \}  =0, 
$
with ${\tau}$ a unitary transformation~\cite{SM}. 
Introducing a virtual copy of the system, which has a flipped Hamiltonian $-H $, the free energy is doubled for a composite system of the two copies, since the real and virtual copies are related by ${\tau}$.
The expansion moments to construct the doubled free energy thus become
\be
f_{n,c} = {\rm Re} \left\{  \Tr [e^{-in \pi (\tilde{H} \otimes \mathds{1} - \mathds{1}\otimes \tilde{H}) } ] \right\} /D^2,
\ee 
which we recognize as the spectra form factor $\mathcal{K}_\mathcal{T} = \Tr[\mathcal{T}]\Tr[\mathcal{T^\dagger}]/D^2$~\cite{RMCircuit}.
The moments can therefore be extracted using the toolbox of randomized measurements as
\be
f_{n,c} \to \overline{(-2)^{-\sum_j s_j}},
\ee
where $s_j$ denotes the measurement outcome of the $j$-th qubit in the computation basis following the quantum state $U_\text{s}^\dagger e^{-i n\pi\tilde{H} } U_\text{s} \ket{0}^{\otimes L}$~\cite{SM}.
Here, $U_\text{s}$ represents one layer of single-qubit gates uniformly sampled from unitary 2-design, and $\overline{\cdots}$ is a dual average over both random unitaries $U_\text{s}$ and quantum shots.
For a local Hamiltonian, the compiling of this quantum circuit solely involves local operations.
This approach thus constitutes as an efficient {\it virtual-copy compiling} protocol, highly compatible with quantum processors restricted to local connectivity.

The other symmetry we consider is a phase-U(1) symmetry.  
This phase-U(1) symmetry is typically present in quantum chemistry problems for the electron particle number conservation, and in a variety of quantum magnetic systems such as Heisenberg and XY models as a conservation of total spin ($S_z$). 
This type of Hamiltonian has a trivial eigen-state, $\ket{ \phi} = \ket{0}^{\otimes L} $, with $S_z = -L$, where we adopt the convention that the computational basis $\ket{0}$ and $\ket{1}$ correspond to $S_z$ eigenvalues being -1 and +1, respectively.
Without loss of generality, we assume its eigen-energy to be zero, which can always be enforced by introducing an overall energy shift. We further define a random product state $\ket{\psi} =\ket{1\psi_2\psi_3\cdots\psi_L}$, where the first qubit always resides in $S_z^{(1)} = 1$ state ensuring orthogonality, $\braket{\phi|\psi}=0$.
We can thus compile QKFE by constructing superposition states $\ket{\pm}=(\ket{\phi} \pm \ket{\psi})/\sqrt{2}\equiv U_\pm\ket{\phi}$, with $U_\pm$ the quantum circuits that prepare these states $\ket{\pm}$~\cite{CiracPRXQ2021,RamilPRXQ2024}.
By measuring the Loschmidt echo probabilities of letting the initial state $\ket{\phi}$ evolve under a local quantum circuit $U_+^\dagger e^{-in\pi\tilde{H}} U_\pm$~\cite{SM}, we obtain the expansion moments
\be
\begin{aligned}
&f_n \rightarrow \overline{P_+ - P_-}, \\
P_\pm \equiv |&\bra{\phi}U_+^\dagger e^{-in\pi\tilde{H}} U_\pm \ket{\phi}|^2,
\end{aligned}
\ee
where the average $\overline{\ldots}$ is taken over different random states $\ket{\psi}$. 
As these probabilities pick up dynamical phases of the state $\ket{\psi}$ accumulated in the Hamiltonian evolution, with $\ket{\phi}$ as a reference state, we refer to this approach as the {\it reference state compiling} protocol for simulating quantum thermodynamics of U(1) symmetric systems.

The essential building block in both  {\it virtual copy} and {\it reference state} compiling protocols is the direct Hamiltonian evolution, 
avoiding large controlled-unitary operations, such that QKFE can be implemented on quantum processors with much less overhead either by constructing trotterized digital quantum circuits, or engineering analogue Hamiltonian evolution.

In the remainder of this letter, we exemplify the feasiblity of our approach by demonstring the above algorithmic quantum simulation protocols on superconducting quantum processors. We design superconducting quantum chips, \emph{Zuchongzhi 2.2}, having $20$ qubits. The architecture design is similar to that previously developed~\cite{wu2021strong}. High fidelity single-qubit and two-qubit gates are achieved through semi-automated calibration protocols. In our experiments, the $R_z (\theta)$ rotations have an infidelity negligible compared to other error sources. The gate fidelity of $R_y (\pi/2)$ exceeds $99.9\%$. The fidelity of CZ gates reaches above {$99.6\%$}. The experimental results are further cross-verified using our most advanced superconducting quantum processor, {\it Zuchongzhi 3.2}, sharing a similar design with {\it Zuchongzhi 3.0} ~\cite{Gao2025}, by selecting $12$ qubits out of the total $107$.

\begin{figure}[!t]
    \centering
    \includegraphics[width=0.5\textwidth]{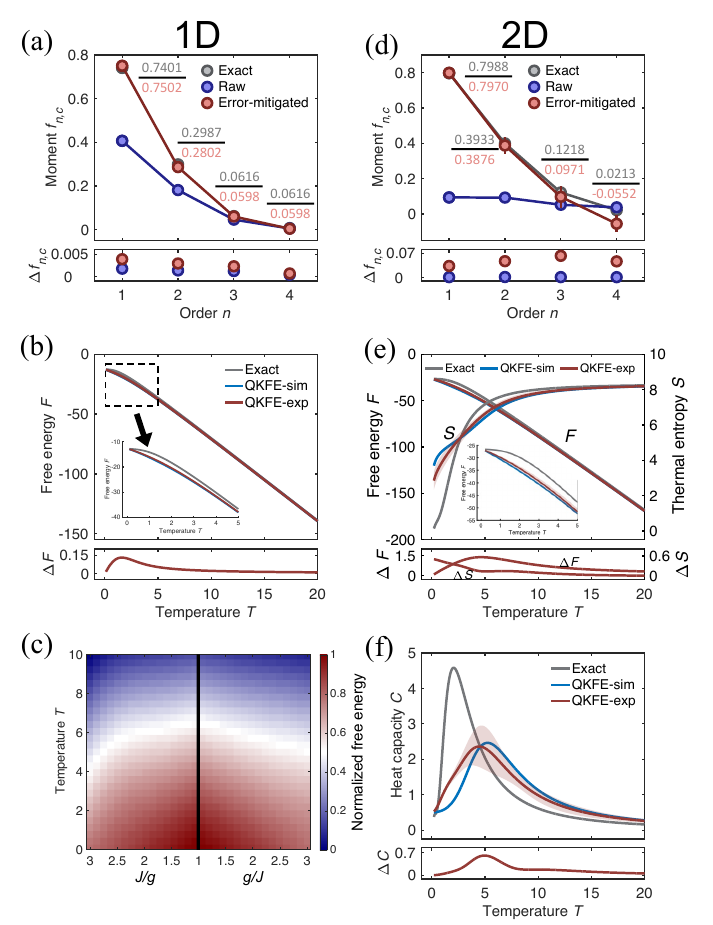}
    \caption{{Digital quantum simulation of TFIM in one- and two-dimensions.
	The left three panels represent  the 1D chain comprising ten qubits with periodic boundary condition,
	and the right panels are for  the two-dimensional $3 \times 4$ lattice.
	The moments ($f_{n,c}$)  in (a) and (d) correspond to the doubled system in the {\it virtual-copy compiling} protocol.   
	The raw and error-mitigated data are compared to the theoretical simulation results (Exact). 
	The deviation of the error-mitigated results from exact values is barely noticeable. 
	(b) and (e), the thermodynamic quantities extracted from the measured moments, for one- and two-dimensional TFIMs, respectively.  
	(c) The observed symmetry between the paramagnetic phase ($g>J$) and ferromagnetic phase ($g<J$) illustrates the experimental observation of  the duality in 1D TFIM\@.
	(f) The heat capacity for the 2D TFIM. 
  	It shows a sizable peak at finite temperature in agreement with the exact results, a signature of the expected phase transition at finite temperature.
	The sampling errors ($\Delta(\cdot)$) of  moments are suppressed down to about $10^{-2}$, by averaging over a sufficiently large number of  quantum shots and random unitaries. We choose $g=J$, and $g=2J$  for the one- and two-dimensional models, respectively. The Ising coupling $J$ is set as an energy unit. 
    }
    }\label{fig:TFIM}
\end{figure}

\begin{figure}[htp]
    \centering
    \includegraphics[width=0.45\textwidth]{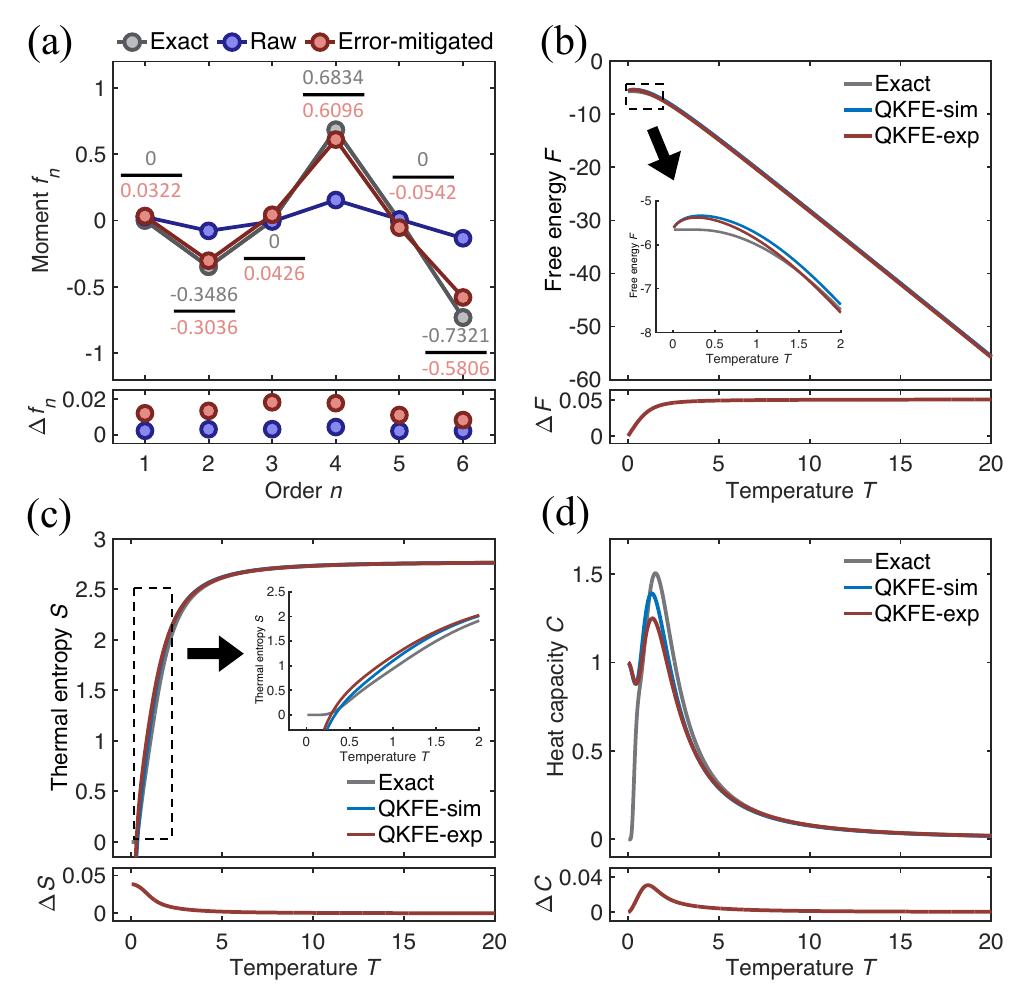}
    \caption{Hybrid digital-analogue quantum simulation of spin XY model.
    (a) the expansion  moments for the free energy\@.  The theoretical (Exact), raw experimental data, and error-mitigated results are provided in the same plot for a comparison. 
    (b) and (c), the free energy and thermal entropy as derived from the experimental measurements. 
    We find excellent agreement for both thermodynamic quantities with the exact results in the entire temperature window.   
	In (d), we provide the heat capacity. The sampling errors ($\Delta(\cdot)$) are below $10^{-2}$. Here, we choose a $2\times 2$ lattice and set the coupling strength $J$ as the energy unit.
    }\label{fig:XYmodel} 
 \end{figure}

\paragraph*{Digital quantum simulations of TFIM.---}
A representative  model for describing quantum magnets is the TFIM
\be 
\label{eq:TFIM} 
H_{\rm Ising} (g, J)= -g \sum_{j=1} ^{L}  X_j 
   - J\sum_{<jk>}  Z_j Z_k,
\ee 
with $J>0$ the Ising ferromagnetic coupling of nearest-neighbor spins, and $g$ the strength of transverse field. 
This model displays a quantum paramagnetic-ferromagnetic duality in one dimension~\cite{SelfDuality2}, and supports a thermodynamic phase transition in two dimensions. 
It has been extensively examined in quantum condensed matter physics and quantum information science, continuously inspiring novel discoveries such as hidden symmetries~\cite{2010_Kiefer_Science}, topological zero modes~\cite{2001_Kitaev_Majorana}, and dynamical phase transitions~\cite{2017_Yao_TimeCrystal,2021_Lukin_Scars}.

TFIM has an anti-commuting symmetry on bipartite lattices~\cite{SM}, enabling the \emph{virtual-copy protocol} for the digital quantum simulations. 
We synthesize the Hamiltonian evolution, $e^{-iH_{\rm Ising} t}$, using Trotterization by sequentially applying the unitary operators $e^{iJZ_jZ_k \delta t}$ and $e^{ig X_j \delta t}$ ~\cite{SM}. 
The entire quantum simulation circuits only involve single qubit rotations, $R_z (\theta)$ and  $R_y(\pm \pi/2)$, and CZ gates~\cite{SM}. 
With these programmable, high-fidelity gates, our superconducting quantum chip accommodates digital quantum simulations of TFIM in both one- and two-dimensions. 
For one-dimension, a qubit ring is selected from the {\it Zuchongzhi 2.2} chip  to simulate a spin chain with a periodic boundary condition, containing up to $10$ qubits. For two-dimensions, the quantum  chip allows for quantum simulations of $2\times 2$, $2\times 3$, {$3\times 3$  and $3\times 4$} lattices with an open boundary condition. We estimate the first four moments ($f_{1,2,3,4}$) with the superconducting quantum processor for one- and two-dimensional TFIMs, as shown in Fig.~\ref{fig:TFIM}{\bf a} and {\bf d}, respectively.

In one dimension, TFIM exhibits a ferromagnetic-paramagnetic duality through the interchange of the coupling parameters, $g$ and $J$~\cite{SelfDuality2}. 
The parameterized Hamiltonians, $H_{\rm Ising} (g, J)$ and $H_{\rm Ising} (J, g)$ have identical energy spectra and thermodynamic properties. 
We determine the free energies in both ferromagnetic and the paramagnetic regime in experiments. 
The error-mitigated moments~\cite{SM} derived from the experimental measurements show excellent agreement with the exact results (Fig.~\ref{fig:TFIM}(a)), with the constructed free energy aligning closely with the exact values across the entire temperature range~(Fig.~\ref{fig:TFIM}(b)). 
The interchange symmetry between the coupling parameters of $g$ and $J$  is clearly demonstrated in the free energy across the regime from $J/g = 3$ to $g/J = 3$ (Fig.~\ref{fig:TFIM}{\bf c}). 
These experimental results showcase the quantitative accuracy of our approach.

In two dimensions, the TFIM  exhibits a second order phase transition at finite temperature, for which the free energy, $F(T)$, develops a singularity in the thermodynamic limit.  This makes the application of the QKFE method more nontrivial than 1D.  
Despite the extra difficulty in simulating the 2D TFIM
due to the larger number of CZ gates required, our error-mitigated~\cite{SM} expansion moments of $3\times4$ lattice still show quantitative agreement with exact results with minimal deviations~(Fig.~\ref{fig:TFIM}(c)). 
The reconstructed free energy also matches the exact results with high accuracy. 
This accuracy enables us to calculate also the thermal entropy and the heat capacity (Fig.~\ref{fig:TFIM}(e),(f)) through taking the first and second derivatives~\cite{SM}. Unlike analogue quantum simulations, where thermodynamic properties have to be measured by preparing the Gibbs ensembles at individual temperatures in experiments, we emphasize that we obtain the thermodynamic quantities in the entire temperature window simultaneously. Consequently, the derivatives are extracted analytically instead of using numerical differences.  
We observe how the heat capacity $C (T)$ develops a peak at finite temperature. The system-size dependence of the heat capacity is provided in Fig.~\ref{fig:Protocal}(d). 
As we increase the system size from $2\times 2$ to $3 \times 4$, we observe the systematic increase of the heat capacity at finite temperature, consistent with its extensive nature. 
The rising peaks of the heat capacity at finite temperature correspond to the expected thermal phase transition of 2D TFIM. 
At zero temperature, the specific heat tends to vanish,
consistent with the gapped ground state.

\paragraph*{Hybrid digital-analogue  quantum simulations of XY model.---}
Finally, we demonstrate that our algorithmic quantum simulation methods are not restricted to digital implementations, but can also be realized through the analogue quantum dynamics of quantum hardware.
The XY model
\be
H_\text{XY}(J) = J\sum_{\langle jk \rangle}(X_j X_k + Y_j Y_k),
\ee
with $J$ the coupling strength between nearest-neighbor spins, the native model of our quantum chip, exhibits a phase-U(1) symmetry. 
To simulate quantum thermodynamics of the XY model, we apply the {\it reference-state compiling} protocol. 
The corresponding quantum circuit consists of layers of digital quantum gates that prepare  and disentangle the superposition states $|\pm\rangle$, with the analogue quantum dynamics  $e^{-iHt}$ sandwiched between these digital layers~\cite{SM}. 
This requires a digital-to-analogue conversion for implementation on the superconducting quantum processor.

To realize the digital-to-analogue conversion, it is crucial to synchronize all qubits to a uniform frequency and calibrate the couplers for  precise qubit-qubit interactions. We use a {multi-qubit excitation propagation} method~\cite{multiqubitexcitation} to calibrate the Hamiltonian evolution parameters. Following the analogue evolution,
{\it cross-entropy benchmarking} is utilized to benchmark the unwanted dynamical phases, which are then compensated by virtual Z-gates~\cite{SM}. With these techniques,  we achieve high-fidelity control over the digital-to-analogue conversion.

The results for the XY model in $2\times2$ lattice are presented in Fig.~\ref{fig:XYmodel}. It is evident that even the raw experimental data for the  $f_n$ moments captures the overall trend as the expansion order $n$ increases. After error mitigation via the zero-noise extrapolation (ZNE)~\cite{SM}, we find quantitative agreement with the exact results.  The free energy and thermal entropy extracted show negligible deviation from the exact values.
The heat capacity, obtained through differentiation as in the fully digital method described above, exhibits a prominent peak at finite temperature, indicative of a thermal phase transition. To conclude, while the present experiment validates our approach on  relatively small system size, we anticipate that this hybrid digital-analogue quantum simulation protocol is readily scalable to larger systems, offering new opportunities to explore superfluid phase transitions and their critical phenomena, such as vortex proliferation in Kosterlitz-Thouless transitions.

\paragraph*{Conclusion.---} 
In summary, we have developed an algorithmic quantum simulation approach for solving complex quantum thermodynamics, and demonstrated it with simulations of the transverse field Ising and XY models on the superconducting quantum processors. 
In these experiments, free energy, local observables, and correlation functions were constructed from the measured spectral form-factors.  
We observed the ferromagnetic-paramagnetic duality of the one-dimensional TFIM, and captured the finite-temperature phase transitions in the two-dimensional transverse-field Ising and XY models, through heat capacity derived from the free energy.
This approach establishes a universal framework for simulating quantum thermodynamics, applicable to both fully digital quantum processors and hybrid digital-analogue platforms. By circumventing the exponential Hilbert space scaling that hinders classical methods, our work offers a promising pathway to tackle complex thermodynamic problems. Ultimately, these results highlight the potential of programmable quantum hardware to unlock previously intractable regimes, with profound implications for both fundamental science and practical applications.

\paragraph*{Acknowledgments.---}
We are indebted to P.~Zoller for insightful suggestions, and acknowledge helpful discussions with A.~del Campo and J.I.~Cirac.
This work is supported by 
Key R \& D Plan of Shandong Province (Grant No. 2024CXPT083), Quantum Science and Technology-National Science and Technology Major Project (2021ZD0300200),
Innovation Program for Quantum Science and Technology of China (Grant No. 2024ZD0300100), 
National Key Research and Development Program of China (Grant No. 2021YFA1400900, Grant No.2024YFB4504002 ), 
National Natural Science Foundation of China (Grant No. 92476203, 11934002, 12404575), 
Shanghai Municipal Science and Technology (Grant No. 2019SHZDZX01, 24DP2600100, 24LZ1400900), 
Natural Science Foundation of Shanghai (Grant No. 23ZR1469600), the Shanghai Sailing Program (Grant No. 23YF1452600) ,
Cultivation Project of Shanghai Research Center for Quantum Sciences (Grant No. LZPY2024),
Anhui Initiative in Quantum Information Technologies,
Special funds from Jinan Science and Technology Bureau and Jinan High Tech Zone Management Committee, the Taishan Scholars Program by Xiaobo Zhu,
Shandong Provincial Natural Science Foundation (Grant No. ZR2022LLZ008), 
the New Cornerstone Science Foundation through the XPLORER PRIZE.
Gang Wu  is supported by the Key-Area Research and Development Program of Guangdong Province (Grant No. 2020B0303060001).
The authors also thank QuantumCTek Co., Ltd., for supporting the fabrication and the maintenance of room-temperature electronics.

\bibliography{references}

\begin{widetext}

\section*{End Matter}

\end{widetext}

\paragraph*{Experimental Setup---} 
The experiments are performed on quantum processor {\it Zuchongzhi 2.2} comprising 20 qubits and {\it Zuchongzhi 3.2} comprising 107 qubits (see~\cite{SM} for layouts of quantum chips), with a design similar to that of our previous device~\cite{wu2021strong}. 
Several upgrades have been implemented to enhance the overall performance of our quantum processors. The center line width and gap width of the qubit capacitor geometry are increased in order to reduce  interface loss, thereby improving the  decoherence time.
Adjacent qubits are now coupled with a smaller direct coupling capacitance. Design parameters are carefully optimized to ensure the presence of a zero-coupling point between neighboring qubits.
As a result, the coupler frequency becomes closer to the optimal point during the  CZ gate operation.
With these improvements in qubit design and advanced control techniques, we achieve high-fidelity quantum gates and analogue quantum evolution for hybrid digital-analogue quantum simulations.

\paragraph*{Error Mitigation---} 
We adopt two error mitigation schemes to mitigate the hardware imperfections, including the GEM and ZNE methods. 
With GEM~\cite{GEM}, we  aim at resolving the isotropic depolarizing error which inevitably occurs in our superconducting quantum processors.  
This error mitigation method is justified when the depolarizing error is dominant, as is true for the relatively large-size and deep quantum circuits.
In the GEM scheme, all error channels are assumed to sum up to an effective depolarizing channel described by 
\be 
    {\rho}_{\rm err}  = (1-p_\text{avg})\rho_{\rm exact}  + p_\text{avg}\frac{I^{\otimes L}}{2^{L}}, 
\ee 
where $\rho_{\rm exact}$ is  the density matrix following a target quantum circuit with no error,  $\rho_{\rm err} $ is the erroneous density matrix, and the average error probability $p_{\rm avg}$ quantifies the overall noise level. 
Taking a Pauli operator $O$, 
its measurement following a erroneous target quantum circuit is 
\be 
\underbrace {\overline{\langle O_{\rm }\rangle_\text{t} }}_{\text{measured}} = (1-p_\text{avg})\underbrace{\langle O \rangle_{\rm t, exact}}_{\text{unknown}} + p_\text{avg}\underbrace{\frac{\Tr[O]}{2^{L}}}_{\text{known}} . 
\ee 
To estimate $p_{\rm avg}$ in experiments, we design an {error estimation circuit}, which suffers almost the same  error channels as the original quantum circuit and at the same time produces an overall identity operation.  
Our compiled quantum circuit for simulating TFIM involves $R_y(\pm \pi/2)$, $R_z(\theta)$ and CZ gates. On our quantum processor, the error rate of the $R_z$ gates is negligible compared to other gates.
We thus flip $R_z(\theta)$ to $R_z(-\theta)$ in certain layers of the quantum circuit in a way that transforms the overall unitary operation to an identity~\cite{SM}.
This is analogous to time-reversing one-half of the quantum circuit. Consequently, we have a erroneous quantum circuit which measures 
\be 
 \underbrace{\overline{\langle O \rangle_\text{e} }}_{\text{measured}} = (1-p_\text{avg})\underbrace{\langle {O}\rangle_{\rm e, exact} }_{\text{known}} + p_\text{avg} \underbrace{\frac{\Tr[O]}{2^{L}}}_{\text{known}}. 
\ee 
Here, $\langle O\rangle_{\rm e,exact} $ is the quantum average with respect to the initial state of the quantum circuit, i.e., $|0\rangle^{\otimes L}$, which is exactly known.
The average error probability $p_{\rm avg}$ is then extracted.

We also apply linear ZNE (LZNE) method for error mitigation~\cite{LZNE}, which is relatively more costly than the GEM approach. 
In ZNE, the noise strength is deliberately amplified, to extrapolate to the zero noise limit. 
Since the dominant errors across the quantum circuit are produced by CZ gates, we thus amplify the noise level by replacing each CZ gate by an odd number ($r$) of CZ gates. 
The noise amplification procedure is performed for both the target quantum circuit and the error estimation circuit (as introduced above in describing GEM). 
For the target quantum circuit and the error estimation circuit, the measurement of a Pauli observable $O$ is denoted as 
$
\overline{\langle O \rangle_\text{t} } |_r
$
and 
$\overline{\langle O \rangle_\text{e} } |_r$, respectively. 
We introduce a ratio 
$
\zeta (r) = \overline{\langle O \rangle_\text{t} } |_r \times \langle O \rangle_{\rm e,exact}   /\overline{\langle O \rangle_\text{e} } |_r . 
$
The error mitigated result for the observable $O$ is given through the linear extrapolation as
\be
\langle O \rangle_{\rm LZNE}  = \frac{1}{2}\left[ 3\zeta (1) -\zeta(3)  \right ]. 
\ee 
A detailed derivation and experimental details about the error mitigation we used in this paper, is discussed in supplementary materials~\cite{SM}.

\end{document}


\title{Supplementary Materials of \\Algorithmic  Quantum Simulations of Quantum Thermodynamics}

\newcommand{\affilUSTCHeFei}{Hefei National Research Center for Physical Sciences at the Microscale and School of Physical Sciences, University of Science and Technology of China, Hefei 230026, China}
\newcommand{\affilUSTCShanghai}{Shanghai Research Center for Quantum Science and CAS Center for Excellence in Quantum Information and Quantum Physics, University of Science and Technology of China, Shanghai 201315, China}
\newcommand{\affilFudan}{State Key Laboratory of Surface Physics, Institute of Nanoelectronics and Quantum Computing, and Department of Physics, Fudan University, Shanghai 200433, China}
\newcommand{\affilQiZhi}{Shanghai Qi Zhi Institute, AI Tower, Xuhui District, Shanghai 200232, China}
\newcommand{\affilShanghaiQuantum}{Shanghai Research Center for Quantum Sciences, Shanghai 201315, China}
\newcommand{\affilInnsbruckTheory}{Institute for Theoretical Physics, University of Innsbruck, 6020 Innsbruck, Austria}
\newcommand{\affilInnsbruckIQOQI}{Institute for Quantum Optics and Quantum Information of the Austrian Academy of Sciences, 6020 Innsbruck, Austria}
\newcommand{\affilHefeiNatLab}{Hefei National Laboratory, University of Science and Technology of China, Hefei 230088, China}
\newcommand{\affilJinan}{Jinan Institute of Quantum Technology and Hefei National Laboratory Jinan Branch, Jinan 250101, China}

\author{Yangsen Ye}
\altaffiliation{These authors contributed equally to this work.}
\affiliation{\affilUSTCHeFei}
\affiliation{\affilUSTCShanghai}

\author{Jue Nan}
\altaffiliation{These authors contributed equally to this work.}
\affiliation{\affilFudan}
\affiliation{\affilQiZhi}
\affiliation{\affilShanghaiQuantum}

\author{Dong Chen}
\altaffiliation{These authors contributed equally to this work.}
\affiliation{\affilFudan}
\affiliation{\affilQiZhi}
\affiliation{\affilShanghaiQuantum}

\author{Torsten V. Zache}
\affiliation{\affilInnsbruckTheory}
\affiliation{\affilInnsbruckIQOQI}

\author{Qingling Zhu}
\affiliation{\affilUSTCShanghai}
\affiliation{\affilHefeiNatLab}

\author{Yiming Zhang}
\affiliation{\affilUSTCHeFei}
\affiliation{\affilUSTCShanghai}

\author{Yuan Li}
\affiliation{\affilUSTCHeFei}
\affiliation{\affilUSTCShanghai}

\author{Xiawei Chen}
\affiliation{\affilUSTCShanghai}

\author{Chong Ying}
\affiliation{\affilUSTCShanghai}
\affiliation{\affilHefeiNatLab}

\author{Chen Zha}
\affiliation{\affilUSTCShanghai}
\affiliation{\affilHefeiNatLab}

\author{Sirui Cao}
\affiliation{\affilUSTCHeFei}
\affiliation{\affilUSTCShanghai}

\author{Shaowei Li}
\affiliation{\affilUSTCShanghai}
\affiliation{\affilHefeiNatLab}

\author{Shaojun Guo}
\affiliation{\affilUSTCHeFei}
\affiliation{\affilUSTCShanghai}

\author{Haoran Qian}
\affiliation{\affilUSTCHeFei}
\affiliation{\affilUSTCShanghai}

\author{Hao Rong}
\affiliation{\affilUSTCHeFei}
\affiliation{\affilUSTCShanghai}

\author{Yulin Wu}
\affiliation{\affilUSTCHeFei}
\affiliation{\affilUSTCShanghai}

\author{Kai Yan}
\affiliation{\affilUSTCShanghai}
\affiliation{\affilHefeiNatLab}

\author{Feifan Su}
\affiliation{\affilUSTCShanghai}
\affiliation{\affilHefeiNatLab}

\author{Hui Deng}
\affiliation{\affilUSTCHeFei}
\affiliation{\affilUSTCShanghai}
\affiliation{\affilHefeiNatLab}

\author{Yu Xu}
\affiliation{\affilUSTCShanghai}
\affiliation{\affilHefeiNatLab}

\author{Jin Lin}
\affiliation{\affilUSTCShanghai}
\affiliation{\affilHefeiNatLab}

\author{Ming Gong}
\affiliation{\affilUSTCHeFei}
\affiliation{\affilUSTCShanghai}
\affiliation{\affilHefeiNatLab}

\author{Fusheng Chen}
\affiliation{\affilUSTCShanghai}
\affiliation{\affilHefeiNatLab}

\author{Gang Wu}
\affiliation{\affilUSTCHeFei}
\affiliation{\affilHefeiNatLab}

\author{Yong-Heng Huo}
\affiliation{\affilUSTCHeFei}
\affiliation{\affilUSTCShanghai}
\affiliation{\affilHefeiNatLab}

\author{Chao-Yang Lu}
\affiliation{\affilUSTCHeFei}
\affiliation{\affilUSTCShanghai}
\affiliation{\affilHefeiNatLab}

\author{Cheng-Zhi Peng}
\affiliation{\affilUSTCHeFei}
\affiliation{\affilUSTCShanghai}
\affiliation{\affilHefeiNatLab}

\author{Xiaobo Zhu}
\affiliation{\affilUSTCHeFei}
\affiliation{\affilUSTCShanghai}
\affiliation{\affilHefeiNatLab}
\affiliation{\affilJinan}

\author{Xiaopeng Li}
\email{xiaopeng_li@fudan.edu.cn}
\affiliation{\affilFudan}
\affiliation{\affilQiZhi}
\affiliation{\affilShanghaiQuantum}
\affiliation{\affilHefeiNatLab}

\author{Jian-Wei Pan}
\email{pan@ustc.edu.cn}
\affiliation{\affilUSTCHeFei}
\affiliation{\affilUSTCShanghai}
\affiliation{\affilHefeiNatLab}

\maketitle

\tableofcontents

\newpage
\section{Methods}

\subsection{The quantum kernel function expansion algorithm} 
QKFE is a digital quantum algorithm for solving thermodynamic problems. 
Its development has been inspired by the classical Kernel Polynomial Method (KPM)~\cite{QKFE,RevModPhys.78.275}.
For a quantum many-body system with Hamiltonian ${H}$, QKFE starts by normalizing the Hamiltonian, 
$
\tilde{H} = ({H} - E_{\text{min}}\mathds{1})/W,  
$
with $W$ and $E_{\rm min}$ the estimates of spectra width, and ground state energy. 
These quantities do not have to be precisely determined. 
In QKFE, the density of states (DOS) is given by 
\be 
\rho(\epsilon) = f_0+  2 \sum_{n=1}^{N} f_n\cos(n\pi\epsilon), 
\ee 
with $\epsilon\in [0, 1]$.
The free energy is then obtained as 

\bea 
\label{eq:freeE} 
&&  F(T, L) = - k_BT\ln 
\left[ 
	  D \sum_{n>0}  f_n \varphi_n (\beta W)   
\right] + E_\text{min},  \\ 
&&  \varphi_0 (x) = \frac{1-e^{-x} }{x}, \,\,\, 
\varphi_n (x) = \frac{2 \left[1-(-1)^n e^{-x}\right] }{x + n^2 \pi^2/x  }  \nonumber 
\eea 
with $\beta$ the inverse temperature, $x = W/k_B T$,  and $D$ the Hilbert space dimension~\cite{QKFE}.
The partition partition function is given by $Z = e^{-\beta F}$.

Physical observables 
${\cal O} (T)$
with respect to Gibbs thermal ensemble 
are obtained through 
\be
\label{eq:observableexpansion} 
{\cal O} (T) = \frac{D}{Z} \left[ \sum_{n>0} d_n \varphi_n(\beta W) \right] e^{-\beta E_\text{min}}
\ee
with $d_n =  \text{Re}\left\{\Tr\left[{O}e^{-in\pi\tilde{{H}} }\right]\right\}/D $. 
In practical computation, the expansion cutoff order $N$ is finite.  
To damp out the the Gibbs oscillations caused by the finite cutoff, the expansion coefficients are corrected by multiplying a Jackson kernel~\cite{QKFE}. 
The kernel function expansion has a uniform convergence with a bounded error, $\omega_\kappa (1/N)$, with  $\omega_\kappa (\delta)$ defined as  $\text{max}|\kappa (x)-\kappa (y)|_{|x-y|\leq\delta}$.

\subsection{The \emph{virtual-copy compiling} protocol} 
The QKFE-based quantum simulations of thermodynamics can be simplified in presence of an anti-commuting symmetry,   $\left\{ {\tau}, H \right\} =0$ with $\tau$ a unitary operator.
We introduce a composite system described by a Hamiltonian ($H_{\rm com}$) involving real and virtual copies, with 
$H_{\rm com} = H\otimes \mathds{1} - \mathds{1}\otimes H$.  
The expansion moments for the free energy of the composite system become 
\be 
f_{n, c} \to \frac{1}{D^2}  \Tr \left[ e^{in\pi \tilde{H}} \right]  \Tr  \left[ e^{-in\pi \tilde{H} } \right]. 
\ee 
The free energy of the composite system is then given by the expansion in Eq.~\eqref{eq:freeE}.
The free energy of the original system is one half of the composite system. 
The corresponding expansion moments to expand the observable $\langle O\otimes O^\dag \rangle$ on the composite system become 
\be 
\label{eq:compoeff} 
d_{n, c} \to  \frac{1}{D^2}  \Tr \left[ O e^{in\pi \tilde{H}} \right]  \Tr  \left[ O^\dag  e^{-in\pi \tilde{H} } \right]. 
\ee 
The composite observable  is then given by the expansion in Eq.~\eqref{eq:observableexpansion}. 
We consider Pauli operators that either commute or anticommute with $\tau$. 
With this property, the observable on a single-copy is given by $\sqrt{ |\langle O\otimes O^\dag \rangle|}$.
One caveat is the sign of the observable cannot be determined this way.

One benefit of introducing the virtual copy is that the measurement of expansion moments becomes much more hardware efficient. 
A spectral form-factor 
$
K_{{\cal T} } \equiv \frac{1}{D^2} \Tr [ {\cal T} ] \Tr[\cal{T} ^\dag ] 
$ 
is given by  averaging   
$\overline{ (-2) ^ {-\sum_j s_j } }$~\cite{{RMCircuit}}, where $s_j$ represents the measurement outcome of $j$-th qubit in the computation basis following the quantum state 
$
 U_{s} ^\dag {\cal T}  U_s|0\rangle^{\otimes L}. 
$
Here, $U_s$ represents a local random unitary, and  $\overline{\ldots}$ is an average over both random unitaries and different quantum shots. 
The local random unitary consists of one layer of single-qubit gates uniformly sampled from unitary 2-design, such as single-qubit Clifford group.  
The \emph{virtual copy compiling} approach avoids the experimental difficulty of engineering large control unitaries.
For a local Hamiltonian, the compiling of this quantum circuit only involves local gate operations or local interactions, accessible to our superconducting quantum processor. 
This compiling protocol  is expected to applicable to other NISQ devices as well, such as trapped ions and neutral atoms.

\subsection{The \emph{reference-state compiling} protocol} 
In presence of phase-$U(1)$ symmetry, or equivalently total spin $S_z$ conservation, compiling the QKFE-based quantum simulations can be simplified by introducing a reference state. 
This symmetry is generally present in quantum chemistry problems for the electron particle number conservation, and in a variety of quantum magnetic systems, e.g., Heisenberg and XY models. 
Such systems have a trivial eigen-state, $\ket{ \phi} $, with $S_z = -L$.  
Without loss of generality, we assume its eigen-energy to be zero, which can always be enforced by introducing a constant shift in the energy.
This zero-energy state is used as the reference state. 
The {\it reference state compiling} protocol starts by preparing GHZ states
$
	\ket{\pm} \equiv \frac{1}{\sqrt{2}} (\ket{\phi} \pm \ket{\psi}) \equiv U_\pm \ket{\phi}, 
$
with $\ket{\psi}$ a random state, that is orthogonal to $\ket{\phi}$~\cite{RamilPRXQ2024,CiracPRXQ2021}. 
We take a random product state 
$\ket{1\psi_1 \psi_2 \cdots \psi_{L-1}}$ for $\ket{\psi}$, 
with the first qubit always in $S_z=1$ state to satisfy the orthogonality. By measuring the  two probabilities following a quantum circuit $U_\pm^\dagger e^{-in\pi\tilde{H}} U_+$ acting on the initial state $\ket{0}^{\otimes L}$,
$
	P_{\pm} \equiv \left|\bra{0}^{\otimes L} U_\pm^\dagger e^{-in\pi \tilde{H}} U_+ \ket{0}^{\otimes L}\right|^2,
$
we obtain the expansion moment $f_n$ as,
\begin{equation}
 f_n \to  \overline{P_+ - P_-} 
\end{equation}
where $\overline{\cdots}$ denotes averaging over sampling random product states.

\subsection{The anti-commutating symmetry of Ising models on bipartite lattices} 
For bipartite lattices, the lattice sites can be divided into $A$ and $B$ sub-groups in a way such that 
each site is connected to sites of the other sub-groups. 
The TFIM defined on bipartite lattices anti-commutes with a unitary operator 
\be 
\tau = \prod_{j\in A, k\in B} Z_j Y_k. 
\ee 
It follows that 
$ \left\{ \tau, X_j \right\}=0$, 
$\left\{ \tau, Z_j Z_k | _{j\in A, k\in B} \right\}=0$. Then we have an anti-commutating symmetry for the Hamiltonian, $\left\{ \tau, H\right\} = 0$. 
This implies a symmetric energy spectrum—given an eigenstate $|E\rangle$ with eigenvalue $E$,  the quantum state $\tau |E\rangle$ has to be  another eigenstate with eigenvalue $-E$. This ensures that the DOS is centro-symmetric, and that the systems described by $H$ and $-H$ have identical free energies.

\newpage
\section{Extended Figures}

\begin{figure*}[htp]
   \centering
   \includegraphics[width=0.9\textwidth]{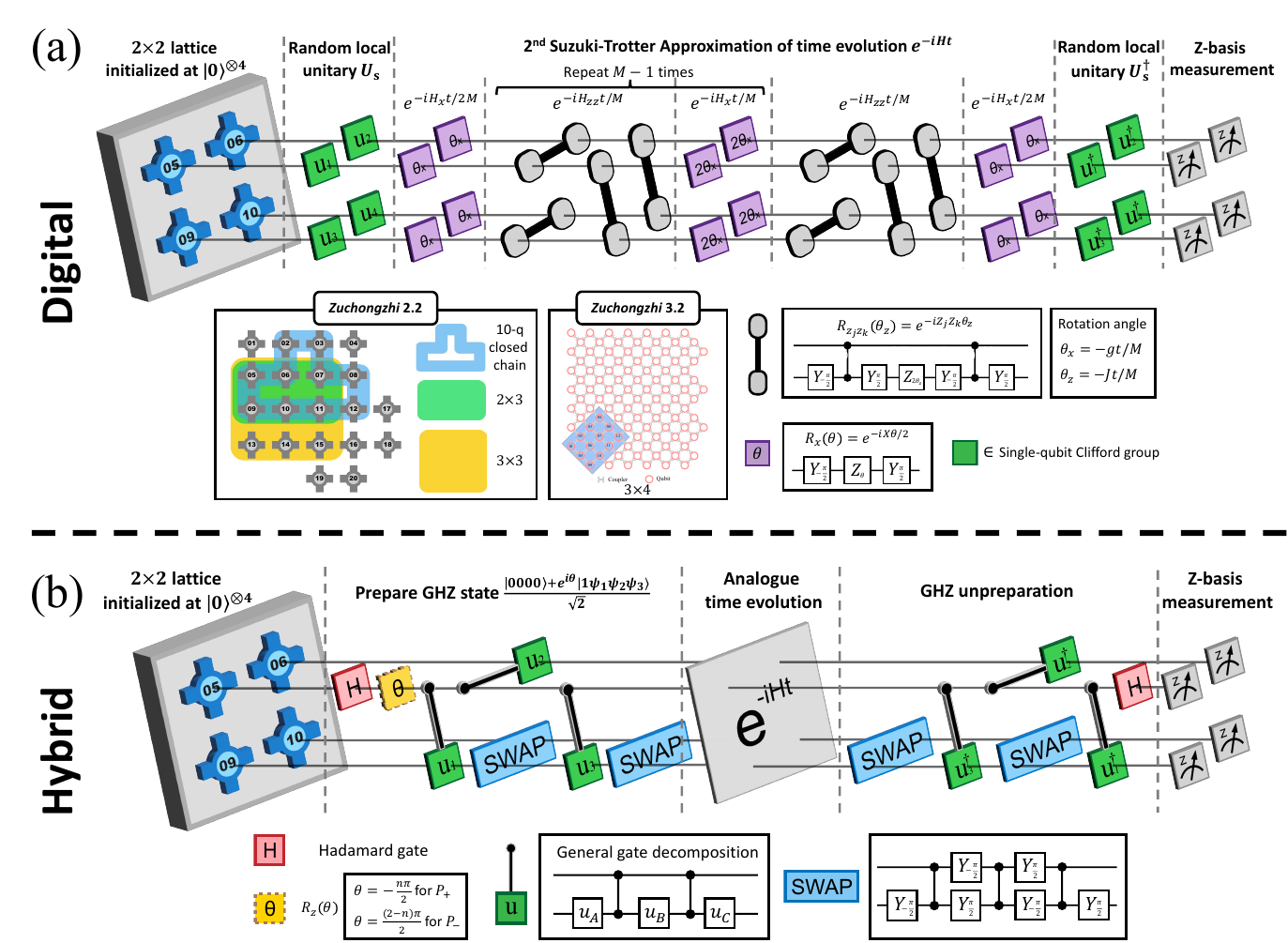}
   \caption{Quantum circuits for hybrid digital-analogue quantum simulation of quantum thermodynamics.
   (a) The digital quantum circuit for measuring expansion moments $f_{n,c}$ of the Hamiltonian with a centrosymmetric energy spectrum. We carry out digital quantum simulation of TFIM on a one-dimensional lattice of ten sites with periodic boundary condition, and  a two-dimensional lattice with system-sizes $2\times2, 2\times3$, $3\times3$ and $3\times4$. The quantum circuit for simulating the $2\times 2$ lattice is provided in (a) as an illustrative example. (b) The hybrid digital-analogue quantum simulation circuit for a two-dimensional XY model on the superconducting quantum processor.   For the {\it reference-state compiling} protocol, the quantum circuit contains random-GHZ state preparation ($U_\pm$), analogue quantum simulation ($e^{-iHt}$), and an inverse operation ($U_+ ^\dag$). This circuit maps nicely to our quantum chip.}\label{fig:Compile}
\end{figure*}

\begin{figure*}[htp]
	\centering
	\includegraphics[width=1.0\textwidth]{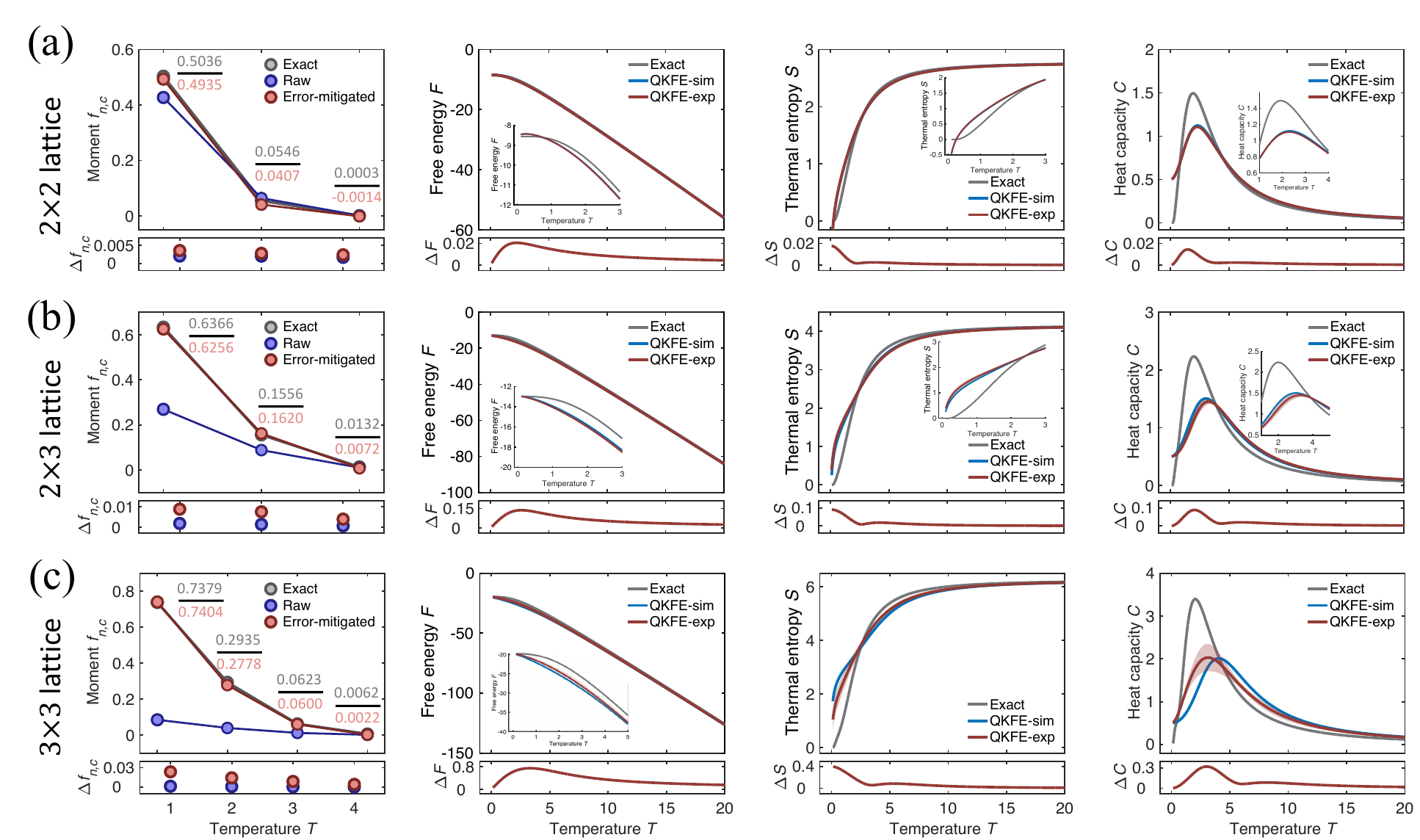}
	\caption{Digital quantum simulation with \emph{virtual-copy protocol} for 2D TFIM.
	(a)(b) The experiment results for $2\times2$, $2\times3$ and $3\times3$ lattices with the LZNE method, respectively. The errorbars ($\Delta(\cdot)$) in this figure are barely noticeable for the standard deviation of moments are suppressed to about $10^{-2}$. Here we choose $g=2J$.
	}\label{fig:TFIM_VirtueCopyProtocol}
\end{figure*}

\begin{figure*}[htp]
	\centering
	\includegraphics[width=1.0\textwidth]{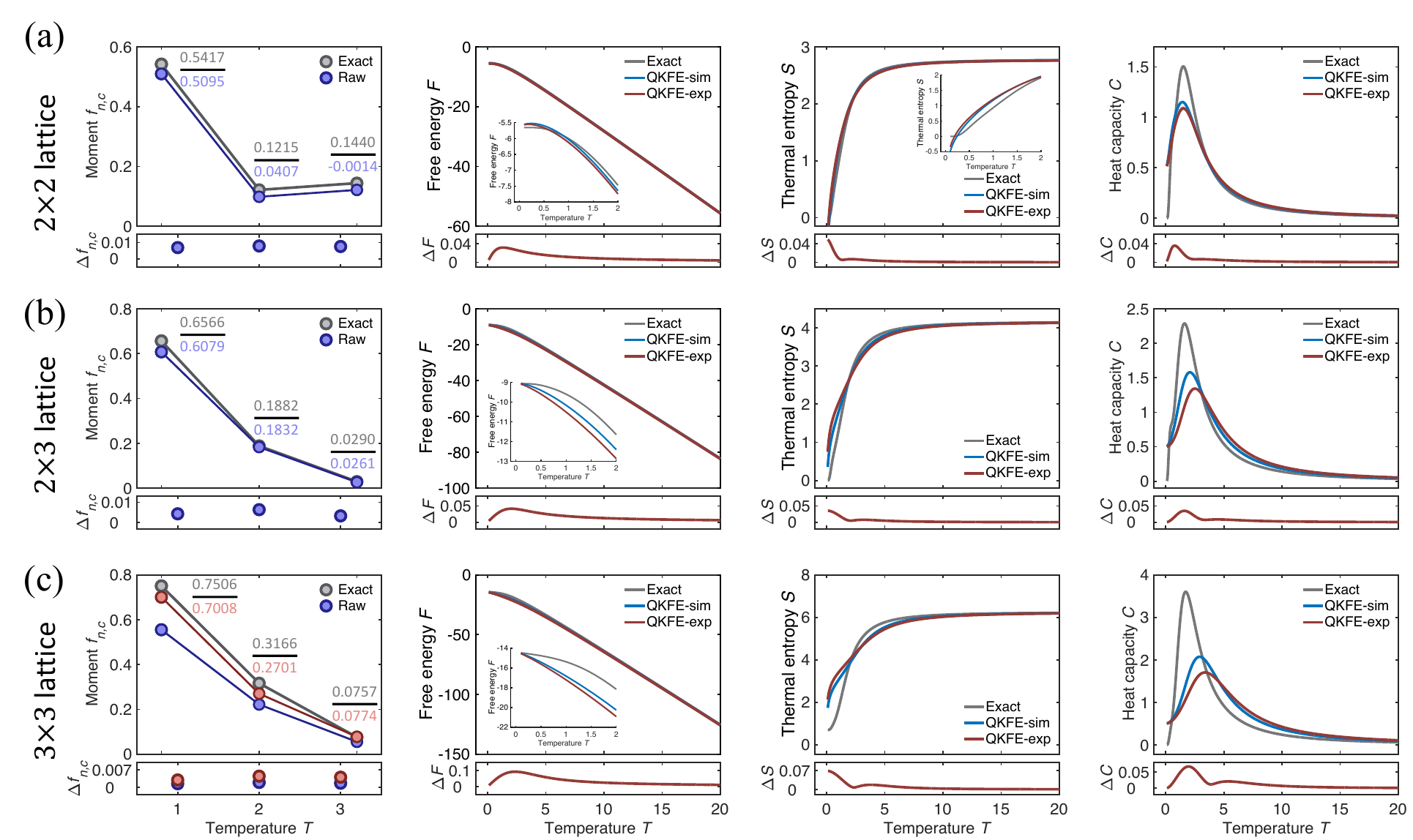}
	\caption{Hybrid quantum simulation with \emph{virtual-copy protocol} for XY model.
	(a)(b)(c) The experiment results for $2\times2$ and $2\times3$ lattices without error mitigation and $3\times3$ lattices with the LZNE method, respectively. The errorbars ($\Delta(\cdot)$) in this figure are also noticeable. The coulpling strength $J$ is set as energy unit.
	}\label{fig:XYmodel_VirtueCopyProtocol}
\end{figure*}

\begin{figure*}[htp]
    \centering
    \includegraphics[width=0.8\textwidth]{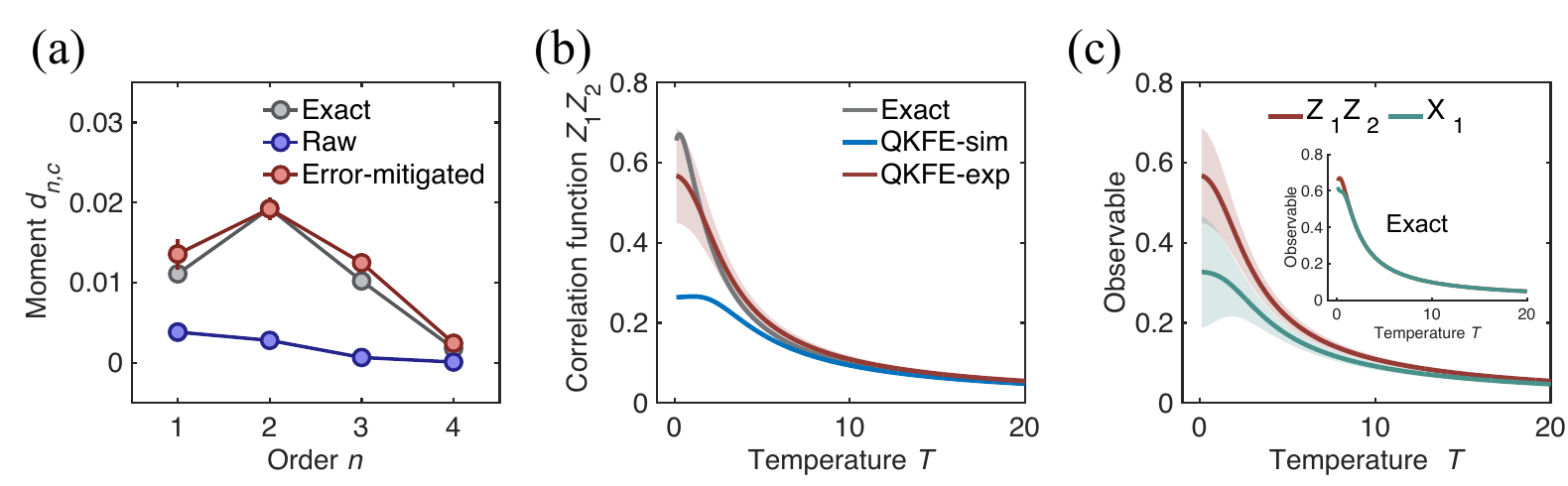}
    \caption{Digital quantum simulation with \emph{virtual-copy protocol} for observables of 1D TFIM.
	The QKFE algorithm is also applicable to measuring the Pauli observables by simply inverting the operators (one layer of single-qubit gates) before or after the time-evolution part $e^{-in\pi\tilde{H}}$. (a) The expansion moments for the correlation function $Z_1 Z_2$ with the GEM method.
	(b) The correlation function $\langle Z_1 Z_2 \rangle$ at finite temperature. 
	(c) The comparison between $\langle Z_1 Z_2 \rangle$ and $\langle X_1 \rangle$.
    The red and green shadows represent the standard deviations estimated by bootstrap. The expansion moments for the correlation function exhibit relatively
	larger disparity from exact results. The discrepancy arises
	from higher quantum shot noise, necessitating extensive statistical averaging and rendering 
	the measurements more susceptible to long-term experimental drifts. Despite this imperfection,
	the digital quantum simulation accurately captures the
	temperature dependence of the correlation. It is worth noting
	that the observed imperfection is not inherent to our digital
	quantum simulation approach, for the results can be further
	refined by improving the long-term stability and collecting a
	larger amount of experimental data. The approaching tendency of two observables with
	escalating temperature, corroborates self-duality of 1D TFIM.
	Here we consider the 1D 10-qubit closed chain at $g=J$.
    }\label{fig:TFIM_Observable}
\end{figure*}

\clearpage
\section{Technial Details}

\subsection{Digital quantum simulation compiling on \textit{Zuchongzhi 2.2} and \textit{3.2}}
Here, we discuss the virtual-copy compiling of TFIM in 1D 10-qubit closed chain, and the 2D lattices of $2\times2$, $2\times3$, $3\times3$ and $3\times4$, on the superconducting processor with digital quantum gates. The TFIM Hamiltonian reads,
\begin{equation}
    \begin{split}
		&H(g,J) = - H_x - H_{zz}, \\
		&H_{zz} = g\sum_{\langle j,k \rangle} Z_j Z_k, \quad H_{x} = J\sum_{j} X_j.
    \end{split}  
\end{equation}
For digital compiling, it is necessary to decompose the evolution operator $e^{-in\pi\tilde{H}}$ into elementary gates. Initially, this decomposition is carried out in $M$ steps, resulting in,
\begin{equation}
    e^{-in\pi\tilde{H}} = e^{-iH(n\pi/W)}e^{in\pi E_\text{min}/W} = {\left(e^{-iH_{x}\tau-iH_{zz}\tau}\right)}^{M}e^{in\pi E_\text{min}/W},
\end{equation}
where $\tau = \frac{n\pi}{MW}$ and $E_\text{min}$ the ground state energy. In the virtual-copy compiling protocol, the directly estimated quantity $\Tr\left[e^{-in\pi\tilde{H}}\right]\Tr\left[e^{in\pi\tilde{H}}\right]/D^2$ is invariant under adding a phase shift, which has thus has been ignored in our circuit decomposition. Given the non-commutativity of $Z_i Z_{i+1}$ and $X_i$, a second-order Suzuki-Trotter decomposition is used for further decomposition,
\begin{equation}
    \begin{split}
        &\text{R.H.S.}\\
		&= e^{-iH_{x}\tau /2} \left(e^{-iH_{zz}\tau} e^{-iH_{x}\tau}\right)^{M-1} e^{-iH_{zz}\tau} e^{-iH_{x}\tau/2} + O\left( M\tau^3 \right),\\
        &= \prod_{j=1} R_{x_j}\left(\theta_x\right) {\left( \prod_{\langle j,k \rangle}  R_{z_j z_k}\left(2\theta_z\right) \prod_{j=1} R_{x_j}\left(2\theta_x\right) \right)}^{M-1} \prod_{\langle j,k \rangle} R_{z_j z_k}\left(2\theta_z\right) \prod_{j=1} R_{x_j}\left(\theta_x\right) + O\left( M\tau^3\right),
    \end{split}  
\end{equation}
with 
\begin{equation}
    \begin{split}
        R_{x_j}(\theta) &\equiv e^{-iX_j\theta/2}, \\
        R_{z_j z_k}(\theta) &\equiv e^{-iZ_j Z_k\theta/2 }, \\
		\theta_x = -g\tau \quad &\text{and} \quad  \theta_z = -J\tau.
    \end{split}
\end{equation}
When measuring a Pauli observable, it is only necessary to insert the operator either before or after $e^{-in\pi\tilde{H}}$.

The feasible unitary operations include single-qubit gates like $R_x(\pi/2)$, $R_x(-\pi/2)$, $R_y(\pi/2)$, $R_y(-\pi/2)$, and $R_z(\theta)$ for any $\theta$, as well as a two-qubit gate, the CZ gate. Therefore, we decompose $R_x(\theta)$ and $R_{z_j z_k}(\theta)$ into $R_z(\theta)$ and other essential gates as
\begin{align}
    R_x(\theta) &= R_y\left(\frac{\pi}{2}\right)\ R_z(\theta)\ R_y\left(-\frac{\pi}{2}\right), \\
    R_{z_{j}z_{k}}(\theta) &= \text{CNOT}^{(j; k)}\ R_{z_k}(\theta)\ \text{CNOT}^{(j; k)}.
\end{align}
Here, ${(j;k)}$ indicates the control (target) qubit as the $j$ ($k$) -th one. A single-qubit unitary rotation can relate the CNOT gate to the CZ gate as
\begin{equation}
    \text{CNOT}^{(j;k)} = R_{y_k}\left(\frac{\pi}{2}\right)\ \text{CZ}^{(j;k)}\ R_{y_k}\left(-\frac{\pi}{2}\right).
\end{equation}

The random unitary $U_\text{s}$ in our circuit is composed of one layer of random single-qubit gates, uniformly sampled from the single-qubit Clifford group, which contains 24 group elements as listed in Table~\ref{24CliffordElement}. As for a Pauli observable, such as the correlation function $Z_jZ_{j+1}$, the random unitary $U_\text{s}$ is extended into layers of single-qubit gates interspersed with one layer of CZ gates to accelerate the convergence of the sampling average.

\begin{table}[htbp]
	\centering
	\caption{\textbf{24 group elements of single-qubit Clifford group}}
	\label{24CliffordElement}
	\setlength{\tabcolsep}{8pt} 
	\small 
	\begin{tabular}{cccc}
		\toprule
		$I$ & $R_x(+\pi/2)$ & $R_x(-\pi/2)$ & $R_x(\pi)$ \\
		$R_y(\pi)$ & $R_x(\pi)\ R_y(\pi)$ & $R_x(+\pi/2)\ R_y(\pi)$ & $R_x(-\pi/2)\ R_y(\pi)$ \\
		$R_y(\pi)\ R_z(-\pi/2)$ & $R_z(-\pi/2)$ & $R_y(+\pi/2)R_z(-\pi/2)$ & $R_y(-\pi/2)R_z(-\pi/2)$ \\
		$R_z(+\pi/2)$ & $R_y(\pi)R_z(+\pi/2)$ & $R_y(-\pi/2)R_z(+\pi/2)$ & $R_y(+\pi/2)R_z(+\pi/2)$ \\
		$R_z(+\pi/2)R_y(-\pi/2)$ & $R_z(-\pi/2)R_y(-\pi/2)$ & $R_y(-\pi/2)$ & $R_z(\pi)R_y(-\pi/2)$ \\
		$R_z(+\pi/2)R_y(+\pi/2)$ & $R_z(-\pi/2)R_y(+\pi/2)$ & $R_y(+\pi/2)$ & $R_z(\pi)R_y(+\pi/2)$ \\
		\bottomrule
	\end{tabular}
\end{table}

\subsection{Hybrid quantum simulation compiling on \textit{Zuchongzhi 2.2}}
\noindent
\emph{Virtual-copy compiling protocol.---}
By simply substituting the gate-compiling evolution for analogue time evolution of superconducting qubits, the measurement outcome following the quantum circuit $U_\text{s}e^{-iHt_n}U_\text{s}^\dagger$ yields the expansion moment $f_\text{n,c}$ of the doubled free energy of the XY model. Since we set the qubit's nearest coupling frequency $J$ to be the energy unit, the evolution time for the $n_\text{th}$ order moment is taken as $t_n = n\pi/JW$.

We firstly scramble the qubits with the local random unitary $U_\text{s}$ at the idle frequencies. Then we synchronize all qubits to a uniform microwave frequency and calibrate  the couplers  for  precise qubit-qubit coupling frequency $J$ to take the analogue time evolution of all qubits. Finally we tune back to the idle frequencies and implement the inverse operation $U_\text{s}^\dagger$ following a phase compensation. Because of the existing rising and falling edges when tuning the qubit-frequency, the effective evolution time (wave time) is a slightly longer than the exact one. We calibrate this using the \textit{multi-qubit excitation propagation} technique. 
In the middle analogue part, the evolution parameters are listed in Table~\ref{tab:EvoParaVC}.

We have performed quantum simulations of the XY model with the system-sizes of $2\times2, 2\times3$ and $3\times3$ in experiment and the results are displayed in Fig.~\ref{fig:XYmodel_VirtueCopyProtocol} in main text.

\begin{table}[htbp]
	\centering
	\begin{tabular}{ccccc}
		\toprule
		\multicolumn{2}{c}{System-size} & $2\times2$ & $2\times3$ & $3\times3$ \\
		\midrule
		\multicolumn{2}{c}{$J\ (2\pi\times\text{MHz})$} & -1.5  & -1.5  & -1.4 \\
		\midrule
		\multirow{3}[2]{*}{Exact time (ns)} & $t_1$ & 58.9  & 36.8  & 24.7 \\
				& $t_2$ & 117.9 & 73.5  & 49.4 \\
				& $t_3$ & 176.8 & 110.3 & 74.1 \\
		\midrule
		\multirow{3}[2]{*}{Wave time (ns)} & $t_1$ & 65    & 42    & 32 \\
				& $t_2$ & 123.5 & 78.5  & 57 \\
				& $t_3$ & 182.5 & 115.5 & 82 \\
		\bottomrule
	\end{tabular}%
	\caption{\textbf{Evolution parameters for virtual-copy compiling of XY model}}
	\label{tab:EvoParaVC}
  \end{table}%

\noindent
\emph{Reference-state compiling protocol.---} 
The quantum circuit of this protocol is also in the digital-analogue-digital configuration, but the local random unitary $U_s$ and its Hermitian conjugate $U_s^\dagger$ are replaced by the GHZ state creation $U_\pm$ and the inverse operation $U_+^\dagger$, respectively. The measurement outcome of this circuit gives $\text{Re}\left\{\Tr[e^{-in\pi\tilde{H}}]\right\}/D = \text{Re}\left\{\Tr[e^{-iHt_n}] e^{-in\pi E_\text{min}/W}\right\}/D$, where the phase shift is nontrivial. This phase can be added by implementing a Z-rotation on the control qubit during the GHZ preparation. For the XY model with centrosymmetric spectrum relative to zero, where $E_\text{min}=-W/2$, the Z-rotation angle is $-n\pi/2$. The evolution 	parameters are listed in Table~\ref{tab:EvoParaRS}.

\begin{table}[htbp]
	\centering
	\begin{tabular}{rcccc}
	  \toprule
	  \multicolumn{3}{c}{System-size} & \multicolumn{2}{c}{$2\times2$} \\
	  \midrule
	  \multicolumn{3}{c}{$J\ (2\pi\times\text{MHz})$} & \multicolumn{2}{c}{-3} \\
	  \midrule
          & \multicolumn{2}{c}{Exact time (ns)} & \multicolumn{2}{c}{Wave time (ns)} \\
	  \midrule
	  $t_1$     & \multicolumn{2}{c}{29.5} & \multicolumn{2}{c}{33.5} \\
	  $t_2$     & \multicolumn{2}{c}{58.9} & \multicolumn{2}{c}{59} \\
	  $t_3$     & \multicolumn{2}{c}{88.4} & \multicolumn{2}{c}{92.5} \\
	  $t_4$     & \multicolumn{2}{c}{117.9} & \multicolumn{2}{c}{122} \\
	  $t_5$     & \multicolumn{2}{c}{147.3} & \multicolumn{2}{c}{151.5} \\
	  $t_6$     & \multicolumn{2}{c}{176.8} & \multicolumn{2}{c}{181} \\
	  \bottomrule
	\end{tabular}
	\caption{\textbf{Evolution parameters for reference state compiling of XY model}}
    \label{tab:EvoParaRS}
  \end{table}

\subsection{Error mitigation}\label{sec:ErrorMitigation}
To ensure the precision of moments measured from quantum circuit, we must tackle the challenge of errors occurring in physical operations due to unwanted or imperfect interactions.
However in the era of noisy intermediate-scale quantum computers, we
can only mitigate the error rather than eliminate it
In the past decades, various methods have been proposed for quantum error mitigation, among which Probabilistic Error Cancellation (PEC)~\cite{PECexperiment} and Zero-Noise Extrapolation (ZNE)~\cite{ZNE1experiment} have been mostly adopted in experiments.
In PEC, a representative error model is firstly learned from device, and then effectively cancelled by sampling Pauli gates from a distribution of noise channel related to the learned model and inverting into corresponding locations in the circuit, which is somewhat similar to error correction.
In contrast, ZNE requires the controlled amplification of the error by an estimated gain factor to extrapolate the ideal zero-noise 
result.
In addition, a newly developed method, called learning-based error mitigation, acquires the information of error model through adjusting the probability distribution of inverted Pauli gates to minimize the difference between ideal value and
erroneous result~\cite{LearningBasedEM}, rather than directly measuring it in naive PEC\@.

However, the above three popular methods all require extremely high measurement cost to learn the error model, and are complicated for experimentalists when implementing a specific quantum algorithm. Furthermore, the prolonged measurement procedure can exacerbate the drifting of hardware performance, which may degrade the experiment results.

Here, we will primarily introduce several main types of error occurring in superconducting quantum computer~\cite{GoogleSurfaceCode}, and then discuss two kinds of error mitigation methods used in our experiment: {Global Error Mitigation} (GEM) 
and {Linear Zero Noise Extrapolation} (LZNE).

\emph{Error channels.---} Here, we give a brief overview of the noise channels commonly 
present in the superconducting qubit systems.
\begin{itemize}
    \item \textbf{Decoherence:} Superconducting qubits are prone to decoherence, a
    phenomenon stemming from the interaction between qubits and their surrounding environment. This interaction 
    results in the loss of quantum information. Decoherence in superconducting 
    qubits manifests in two primary forms: energy relaxation and dephasing. Energy 
    relaxation occurs when the qubit dissipates energy into the environment, 
    primarily due to its coupling to lossy defects in the substrate. Dephasing, on 
    the other hand, involves the loss of phase coherence in the qubit, primarily 
    induced by frequency fluctuations stemming from white noise and 1/\textit{f} 
    noise in the environment.

    \item \textbf{Control errors:} During the calibration and execution of quantum 
    gates, coherent errors known as control errors may arise. These errors are primarily attributed to 
    imperfections in the control pulses, such as amplitude and phase deviations, 
    leading to deviations from the intended operation. Over time, parameter drifts 
    in quantum hardware and control electronics can further contribute to these 
    errors. Additionally, unintentional interactions resulting from microwave 
    crosstalk can also contribute to these errors.

    \item \textbf{Readout errors:} Readout errors in superconducting qubits consist primarily of two components: 
    separation errors and incoherent transitions during measurements. Separation 
    errors occur when the states of the qubit are not sufficiently distinct in the 
    IQ plane, resulting in misidentification of the qubit's state. Incoherent 
    transitions, on the other hand, can be triggered by thermal excitation or 
    energy relaxation of the qubit during the readout process, ultimately leading 
    to inaccurate measurement outcomes. In this experiment, a transition 
    matrix which represents the transition probability from the measurement state 
    to ideal state is used to mitigate the readout error. 

    \item \textbf{Leakage errors:} In quantum computation, qubit represents a  
    two-level system. However, the weak anharmonicity exhibited by the transmon 
    qubit, coupled with the operation mechanism of the CZ gate, can result in the 
    excitation of the qubit to higher energy levels during gate operations. This 
    undesired population transfer in the higher energy levels is referred to as 
    leakage errors. Leakage errors have the potential to propagate throughout the 
    quantum circuit, thereby compromising the fidelity of the quantum computation. Several factors, 
    including thermal excitations and imprecise control of CZ gates, can contribute 
    to leakage errors. 

\end{itemize}

Although the aforementioned models effectively capture a significant portion of 
errors encountered in the experimental setup, they tend to underestimate the 
experimentally calibrated decay of fidelity observed in benchmarking 
experiments. Several potential factors contribute to the observed discrepancy 
between the implemented error models and the benchmarking outcomes. 
These could 
include distortions in flux control, the coupler 
transitioning to its first excited state~\cite{sung2020realization}, drifts 
in TLS frequencies, and instabilities in control electronics.

\emph{Global Error Mitigation.---}
With the global error mitigation (GEM) method~\cite{GEM}, we  aim at resolving the isotropic depolarizing error which inevitably occurs in all present  quantum hardware including our superconducting quantum chips. 
This error mitigation method is justified when the depolarizing error is dominant, as is true for the relatively large-size and deep quantum circuits
~\cite{GoogleSurfaceCode,PECexperiment,ZNE1experiment}. 
A $n$-qubit isotropic depolarizing channel reads as 
\begin{equation}
    \mathcal{E}^{(n)}(\rho) = (1-p)\rho + \frac{p}{4^n-1} \sum_{\sigma \in \Lambda} \left(\bigotimes\limits_{j=1}^{n}\sigma_j\right) \rho  \left(\bigotimes\limits_{j=1}^{n}\sigma_j\right), \label{eq:31}
\end{equation}
where $\rho$ is the density matrix, $\Lambda = {\{X,Y,Z,I\}}^{\otimes n} \backslash I_{2^n}$ and $p$ is the probability of one Pauli channel occurring. Using the equality
$\sum_{\sigma\in\Lambda} \sigma\rho\sigma = 2^n I_{2^n} - \rho$, we eliminate the pauli channels, and adapt Eq.~\eqref{eq:31} into the following form,
\begin{equation}
    \mathcal{E}^{(n)}(\rho) = (1-p_\text{avg})\rho + p_\text{avg}\frac{I_{2^n}}{2^n},
\end{equation}
with $p_\text{avg}$ the average depolarizing error rate and it holds that $p_\text{avg} = \frac{4^n}{4^n-1}p$.

For the one single-qubit and two-qubit cases, where the total density matrix is partially depolarized, the corresponding channels should be written in the partial trace form,
\begin{gather}
    \mathcal{E}_i(\rho) = (1-p_\text{avg})\rho + p_\text{avg}\Tr_i\left[\rho\right]\otimes\frac{I_2^{(i)}}{2} \label{eq:33},\\
    \mathcal{E}_{i,j}(\rho) = (1-p_\text{avg})\rho + p_\text{avg}\Tr_{i,j}\left[\rho\right]\otimes\frac{I_2^{(i)}}{2}\otimes\frac{I_2^{(j)}}{2} ,
\end{gather}
where $i,j$ denote the index of qubits suffering the depolarizing error.

When dealing with all combinations of depolarizing error channels along with the unitary operations in a quantum circuit, there will be many partial traced terms which still contains
some coherence in the remaining qubits. However, in the GEM method, these terms are replaced by the maximally mixed state and an effective depolarizing channel on the entire quantum state is given as~\cite{GEM}
\begin{equation}
    \rho_\text{err}(t) = (1-p_\text{avg})\rho_\text{exact}(t) + p_\text{avg}\frac{I^{\otimes L}}{2^{L}}\label{eq:45},
\end{equation}
with $\rho_\text{err}(t)$ the density matrix of the final state suffering depolarizing error, $\rho_\text{exact}(t)$ the exact one without error, and $p_\text{avg}$ the {global error rate}
roughly estimating the severity of the plaguing depolarizing error in overall circuit and $L$ the number of qubits. This approximation is reasonable for a quite deep circuit
with many qubits and thus we only apply it to measuring the expansion moments of observables $Z_1Z_2$ and $X_1$ for 10-qubit TFIM, where the circuit depth (number of layers) is ranged from 33 to 89.

We express Eq.~\eqref{eq:45} in operator's expectation value,
\begin{equation}
    \underbrace{\overline{\langle O\rangle}_\text{t}}_{\text{measured}} = (1-p_\text{avg})\underbrace{{\langle O\rangle}_\text{t}}_\text{unknown} + p_\text{avg}\underbrace{\frac{\Tr[ O]}{2^{L}}}_{\text{known}} \label{eq:37},
\end{equation}
where the subscript \quos{t}\quos\ refers to the {target circuit} which is just the QKFE circuit, $\overline{\langle O\rangle}_\text{t} = \Tr[O\rho_\text{err}(t)]$ 
is the experimental result of target circuit with error, $\langle O\rangle_\text{t} = \Tr[ O\rho_\text{exact}(t)]$ is the error-free value.

\begin{figure}[htp]
    \centering
    \includegraphics[width=0.88\textwidth]{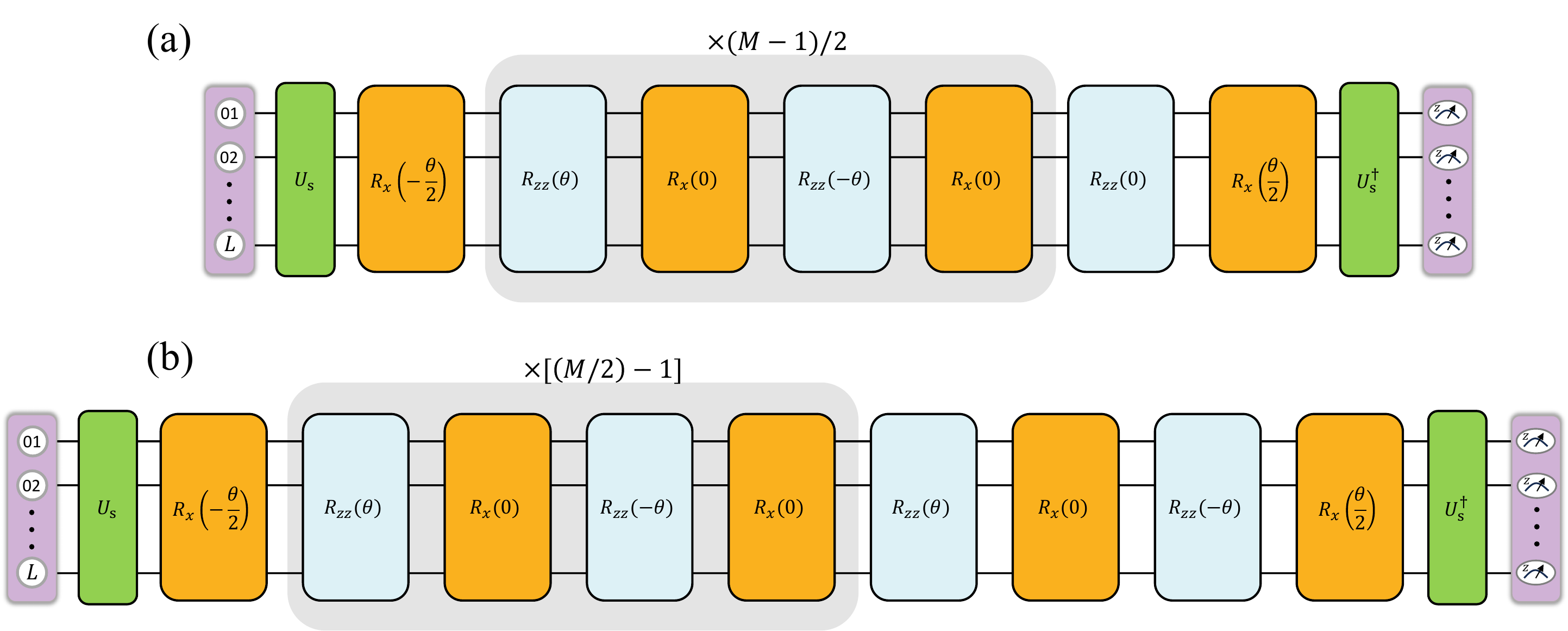}
    \caption{Error estimation circuit for digital simulation of the TFIM.
    (a) The number of time steps $M$ is odd. (b) $M$ is even. Since in our gate decomposition, all rotations of $\theta_x$ and $\theta_z$ are performed on $z$-axis whose error is negligible, we force the ZZ layers and X layers to rotate forward or backward with $\theta$, or simply do not rotate, 
    which will offsets all unitary operations but reserve the error channels as in the target circuit.
    }\label{fig:EEC}
 \end{figure}

To estimate $p_\text{avg}$ in experiment, we design an {error estimation circuit}, which suffers almost simultaneous errors channels compared to target circuit,
nevertheless, with overall effect of unitary operations being unity. Fig.~\ref{fig:EEC} illustrate the error estimation circuit for digital simulation of the TFIM. It is achieved by only adjusting the rotation angle of $R_z(\theta)$ to counteract all unitary operations. This is accounted for by two reasons,
\begin{itemize}
    \item $p_\text{avg}$ is determined by the number and the error rate of all single-qubit and two-qubit gates in a circuit;
    \item The error associated with $R_z$ rotation is negligible compared to other error sources on the superconducting quantum chip, \emph{Zuchongzhi 2.2}.
\end{itemize}
 From the error estimation circuit, we have
\begin{equation}
    \underbrace{\overline{\langle O\rangle}_\text{e}}_{\text{measured}} = (1-p_\text{avg})\underbrace{\langle O\rangle_\text{e}}_{\text{known}} + p_\text{avg}\underbrace{\frac{\Tr[ O]}{2^{L}}}_{\text{known}} \label{eq:38},
\end{equation}
where $\overline{\langle O \rangle}_\text{e} = \Tr[O \rho_\text{err}(t=0)]$ is the experimental result of the error estimation circuit, and $\langle O \rangle_\text{e} = \Tr[O\rho_\text{exact}(t=0)] = \bra{\bm{0^{L}}} O \ket{\bm{0^{L}}}$ is the error-free value on the initial state (easy to estimate and dependent on the operator $O$).

The GEM method is summarized in following procedure:
\begin{enumerate}
    \item Calibrate $p_\text{avg}$ by measuring $\overline{\langle O\rangle}_\text{e}$ right after running the target circuit which yields $\overline{\langle O\rangle}_\text{t}$,
    which ensures the drifting of device's performance is as negligible as possible during these two measurements,
    \begin{equation}
       p_\text{avg} = \frac{ {\langle O \rangle}_\text{e} - \overline{\langle O\rangle}_\text{e}}{ {\langle O \rangle}_\text{e} - \Tr[ O]/2^{L}};
    \end{equation} 
    \item Estimate the error-mitigated expectation values, denoted as $\langle O\rangle_\text{m}$, which is a biased estimator of the error-free value $\langle O\rangle_\text{t}$,
    \begin{equation}
        \langle O\rangle_\text{m} = \frac{\overline{\langle O\rangle_\text{t}} - p_\text{avg}\Tr[O]/2^{L}}{1-p_\text{avg}};
        \label{eq:39}
    \end{equation}
    \item Repeat steps 1$\sim$2 by sampling random unitaries until the error-mitigated average converges.
\end{enumerate}

GEM method is quite user-friendly for experimental implementation, with only twice the original measurement cost. However the precision of error-mitigated result is not as satisfying as that of LZNE, which will be discussed in next subsection. In our experiment, this method 
is only used in measuring ensemble average of local observables which require a large number of sampling times to converge.

\emph{Linear Zero Noise Extrapolation.---}
We also apply linear zero noise extrapolation (LZNE) method for error mitigation~\cite{LZNE}. 
In the original framework of ZNE, it is assumed that the two-qubit gate, such as CZ gate, causes the dominant errors across quantum circuit,  
which is justified by the contrast between error rate of CZ, ranging from $0.2\%\ \text{to}\ 1.0\%$ and that of single-qubit gate, $0.03\%\ \text{to}\ 0.16\%$ . 

\begin{figure}[htp]
    \centering
    \includegraphics[width=0.30\textwidth]{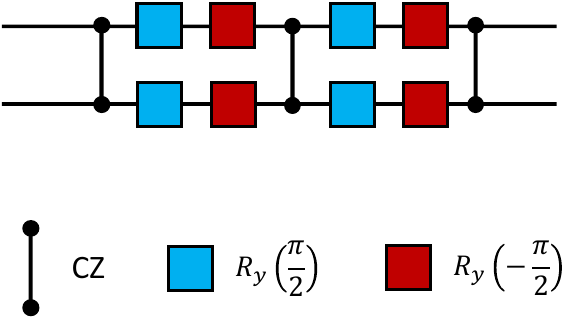}
    \caption{Increasing the error channels of CZ by 3 times.
    The rotation along $y$-axis forth and back is to avoid the pulse distortion of sequential CZ gates.~\cite{GoogleSurfaceCode}
    }\label{fig:6}
\end{figure}

Therefore we replace one single CZ gate with an odd number of repeated CZ gates, which still acts as an equivalent CZ gate, while at the same time amplifies the noise level by multiple times (see the illustration in Fig.~\ref{fig:6}). 
We implement the target circuit and the error estimation circuit, respectively, and extrapolate to the zero noise limit.

If we increase the number of CZ gates by a factor of $r$ in experiment, the equivalent error rate of one CZ gate is scaled up to $1 - {(1-p)}^r$, with $p$ the occurring probability of a two-qubit depolarizing channel.
We denote the expectation value of $\hat{O}$ operator in the target circuit where one single CZ gate is repeated by $r$ times, as $\overline{\langle O\rangle}_\text{t}|_r$ and it holds that
\begin{equation}
    \begin{split}
        &\overline{\langle O\rangle}_\text{t}|_r = \sum_{i=0}^{N_\text{c}} {\left[1-{(1-p)}^r\right]}^i {(1-p)}^{r(N_\text{c}-i)} \langle O\rangle_{\text{t}, i} \\
        &= {(1-p)}^{rN_\text{c}} \sum_{i=0}^{N_\text{c}} {\left[{(1-p)}^{-r}-1\right]}^i \langle O\rangle_{\text{t}, i} \\
        &= \left[1-rN_\text{c}p + \frac{1}{2}rN_\text{c}(rN_\text{c}-1)p^2 + O(p^3)\right] \left[\langle O\rangle_\text{t} + r\langle O\rangle_{\text{t},1}p + \left(r^2\langle O\rangle_{\text{t},2} + \frac{1}{2}r(r+1)\langle O\rangle_{\text{t},1}\right)p^2 + O(p^3)\right] \\
        &= \langle O\rangle_{\text{t}} \left\{1 + \underbrace{\left(r\frac{\langle O\rangle_{\text{t},1}}{\langle O\rangle_{\text{t}}} - rN_\text{c}\right)}_{\alpha_\text{t}}p + \underbrace{\left[\frac{1}{2}rN_\text{c}(rN_\text{c}-1) + \left(\frac{1}{2}r(r+1) - r^2N_\text{c}\right)\frac{\langle O\rangle_{\text{t},1}}{\langle O\rangle_{\text{t}}} + r^2\frac{\langle O\rangle_{\text{t},2}}{\langle O\rangle_{\text{t}}}\right]}_{\beta_\text{t}}p^2\right\} + O(p^3).
    \end{split}
    \label{eq:40}
\end{equation}
The terms expressed explicitly are up to the second order in $p$, where $N_\text{c}$ is the number of CZ gates, $\langle O \rangle_{\text{t},i}$ is the sum of the expectation values over all partially traced density matrices where $i$ CZ gates are affected by an isotropic depolarizing
channel in target circuit. Additionally $\langle O \rangle_{\text{t}}$ is the error-free result of target circuit.

\par
As for the expectation value of $O$ operator in the error estimation circuit, where one single CZ gate is repeated by $r$ times, denoted as $\overline{\langle O \rangle}_\text{e}|_r$, we have
\begin{equation}
    \begin{split}
        &\overline{\langle O \rangle}_\text{e}|_r = \sum_{i=0}^{N_\text{c}} {\left[1-{(1-p)}^r\right]}^i {(1-p)}^{r(N_\text{c}-i)} \langle O \rangle_{\text{e},i} \\
        &= \langle O \rangle_\text{e} \left\{1 + \underbrace{\left(r\frac{\langle O \rangle_{\text{e},1}}{\langle O \rangle_\text{e}} - rN_\text{c}\right)}_{\alpha_\text{e}}p + \underbrace{\left[\frac{1}{2}rN_\text{c}(rN_\text{c}-1) + \left(\frac{1}{2}r(r+1) - r^2N_\text{c}\right)\frac{\langle O \rangle_{\text{e},1}}{\langle O \rangle_\text{e}} + r^2\frac{\langle O \rangle_{\text{e},2}}{\langle O \rangle_\text{e}}\right]}_{\beta_\text{e}}p^2\right\} \\ 
        &\,\,\,\,+ O(p^3),
    \end{split}
    \label{eq:41}
\end{equation}
with $\langle O \rangle_{\text{e},i}$ the sum of the expectation values over all partially traced density matrices where $i$ CZ gates are plagued by isotropic depolarizing channels in the error estimation circuit.

\par
Here we define the ratio of $\langle O \rangle_\text{t}|_r$ and $\langle O \rangle_\text{e}|_r$, as $\zeta(r)$
\begin{equation}
    \begin{split}
        &\zeta(r) = \frac{\overline{\langle O \rangle}_\text{t}|_r}{\overline{\langle O \rangle}_\text{e}|_r} \langle O \rangle_\text{e} \\
        &= \langle O \rangle_{\text{t}} \left[1 + (\alpha_\text{t} - \alpha_\text{e})p + (\beta_\text{t} - \beta_\text{e} + \alpha_\text{e}^2 - \alpha_\text{t}\alpha_\text{e})p^2 + O(p^3)\right]\\
        &= \langle O \rangle_{\text{t}}
        \left\{
            1 + r\left(\frac{\langle O \rangle_{\text{t},1}}{\langle O \rangle_{\text{t}}} - \frac{\langle O \rangle_{\text{e},1}}{\langle O \rangle_\text{e}}\right)p +
            \left[
                \left(\frac{1}{2}r(r+1) - r^2\frac{\langle O \rangle_{\text{e},1}}{\langle O \rangle_\text{e}}\right) 
                \left(\frac{\langle O \rangle_{\text{t},1}}{\langle O \rangle_{\text{t}}} - \frac{\langle O \rangle_{\text{e},1}}{\langle O \rangle_\text{e}}\right) 
                + r^2\left(\frac{\langle O \rangle_{\text{t},2}}{\langle O \rangle_{\text{t}}} - \frac{\langle O \rangle_{\text{e},2}}{\langle O \rangle_\text{e}}\right)                  
            \right]
            p^2 \right. \\ 
& \,\,\,\, \left.            + O(p^3)
        \right\}.
    \end{split}
    \label{eq:43}
\end{equation}
The expectation value after linear zero-noise extrapolation $\langle O \rangle_\text{LZNE}$ is then give by
\begin{equation}
    \begin{split}
        \langle O\rangle_\text{LZNE} &= \frac{3\zeta(1) - \zeta(3)}{2} \\
        &= \langle O \rangle_{\text{t}}
        \left\{
            1 + \frac{3}{2}
            \left[
                \left(2\frac{\langle O \rangle_{\text{e},1}}{\langle O \rangle_\text{e}} - 1\right)
                \left(\frac{\langle O \rangle_{\text{t},1}}{\langle O \rangle_{\text{t}}} - \frac{\langle O \rangle_{\text{e},1}}{\langle O \rangle_\text{e}}\right)
                -2\left(\frac{\langle O \rangle_{\text{t},2}}{\langle O \rangle_{\text{t}}} - \frac{\langle O \rangle_{\text{e},2}}{\langle O \rangle_\text{e}}\right)
            \right]p^2 \right.  \\ 
 &\left.          + O(p^3)
        \right\}.
    \end{split}
    \label{eq:44}
\end{equation}
As our error estimation circuit (illustrated in Fig.~\ref{fig:EEC}) has fairly similar error channels compared to the target circuit, where only some $R_z(\theta)$ gates (whose error are negligible on \textit{Zuchongzhi 2.2})
are reversed or erased, we know that the ratio $\langle O\rangle_{\text{t}, i}/\langle O \rangle_{\text{t}}$ will potentially be close to $\langle O \rangle_{\text{e},i}/\langle O \rangle_\text{e}$, leading to a decrease 
in quadratic deviation of $\langle O\rangle_\text{ZNE}$~\cite{LZNE}. 

\par
In summary, we generalize the procedure of LZNE into three steps:
\begin{enumerate}

    \item Run the target circuit and error estimation circuit at $r=1$ and $r=3$, respectively, which gives the expectation values $\overline{\langle O \rangle}_\text{t}|_1$, $\overline{\langle O \rangle}_\text{t}|_3$, $\overline{\langle O \rangle}_\text{e}|_1$, $\overline{\langle O \rangle}_\text{e}|_3$
    . Then estimate values $\zeta(1)$ and $\zeta(3)$ with Eq.~\eqref{eq:43}.

    \item Extrapolate linearly to the limit of zero error, $r=0$, using $\zeta(1)$, $\zeta(3)$, and then provide a biased estimator, $\langle O\rangle_\text{ZNE}$, of the error-free value $\langle O \rangle_{\text{t}}$.
    
    \item Repeat steps 1$\sim$2 by sampling random unitaries. The sampling average of all expectation values after LZNE  is a biased estimator of the the ideal moment.

\end{enumerate}

\begin{figure}[htp]
    \centering
    \includegraphics[width=0.95\textwidth]{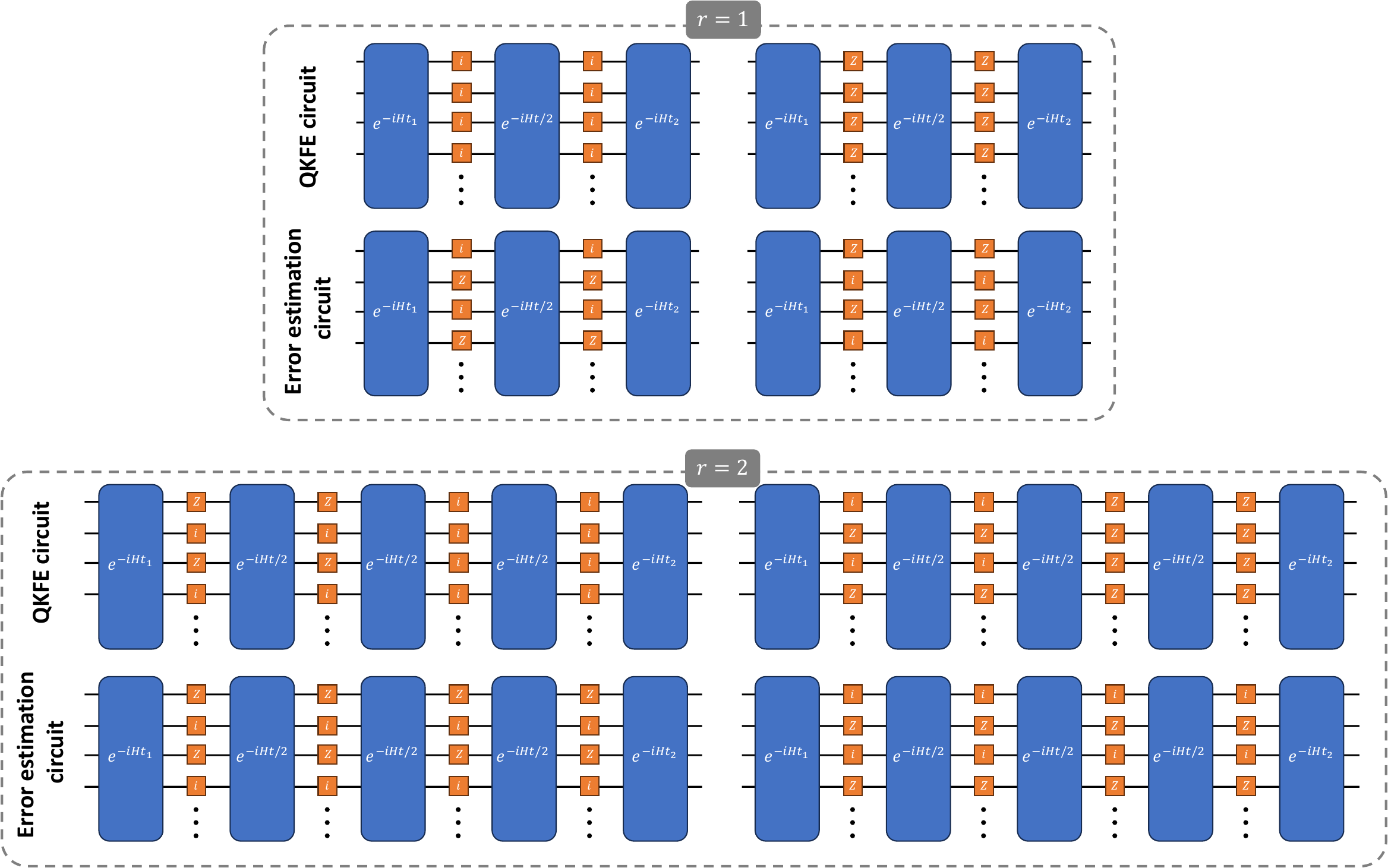}
    \caption{Splitting analogue evolution for $3\times3$ lattice of XY model.
	To use the LZNE method by amplifying noise factor to $r=2$, there are four types of circuits needed to be run, the QKFE circuit and the error estimation circuit at $r=1$ and $r=2$, respectively. The operators $Z_1I_2Z_3I_4\cdots$ and $I_1Z_2I_3Z_4\cdots$ anticommute with the XY Hamiltonian and thus reverse the evolution part stuck in the middle, but the $Z_1Z_2Z_3Z_4\cdots$ commutes with it and causes no effect. The total duration of analogue evolution in the circuit of $r=2$ case is two times longer than that of $r=1$ case and thus has the twice larger noise. Because of the slight difference between the durations of the idle gate and the Z gate, for each type of circuit we further devise two kinds of operators combination to average the effect of this slight difference.
	So there are 16 combinations of the LZNE method with $r=1$ and $r=2$. For each random unitary, we run all these combinations of circuits independently and average their results. The total time duration is $t$, and $t_1 + t_2 =t$.
    }\label{fig:Analogue9qEM_VirtualCopy}
\end{figure}

\emph{Experimental details.---}Here we show crucial details of the error mitigation used in this paper.
\begin{itemize}
	\item \textbf{Digital simulation with virtual-copy protocol:} All Density of States (DOS) moments were mitigated using the Linear Zero-Noise Extrapolation (LZNE) method.
	\begin{itemize}
		\item For smaller systems---the 10-qubit 1D chain, and the $2\times2$ and $2\times3$ TFIM lattices---we set the noise amplification factor to $r=3$ by tripling every CZ gate.
		\item For larger systems, a full amplification of $r=3$ resulted in deviations from the required linear noise response. We therefore implemented a partial amplification scheme. For the $3\times3$ lattice, a subset of 3 (out of 24) CZ gates was tripled, yielding an effective factor of $r=1.25$. For the $3\times4$ lattice, we used an effective factor of $r=21/17$. The final result was averaged over 8 and 4 different subset choices, respectively, to ensure an unbiased estimation.
		\item In contrast, Pauli observable measurements were processed using the Global Error Mitigation (GEM) method.
	\end{itemize}

	\item \textbf{Hybrid simulation with virtual-copy protocol:} This protocol's shallow circuit structure makes it robust against errors.
	\begin{itemize}
		\item For the $2\times2$ and $2\times3$ XY model lattices, the raw data already matched theoretical predictions accurately, rendering error mitigation unnecessary.
		\item For the larger $3\times3$ lattice, we applied LZNE with noise factors $r=1$ and $r=2$. The $r=2$ level was realized by doubling the analog evolution time while preserving the net unitary operation. This was achieved by inserting anticommuting operators to reverse evolution direction in specific segments (see Fig.~\ref{fig:Analogue9qEM_VirtualCopy} for details).
	\end{itemize}

	\item \textbf{Hybrid simulation with reference state protocol:} We applied the LZNE method to the $2\times2$ XY model with noise factors $r=1.5$. To achieve an effective factor of $r=1.5$, we selected a subset of 6 (out of 24) CZ gates, repeated this subset three times, and averaged the final results over 8 distinct subset selections.
\end{itemize}

\begin{figure}[htp]
    \centering
    \includegraphics[width=0.65\textwidth]{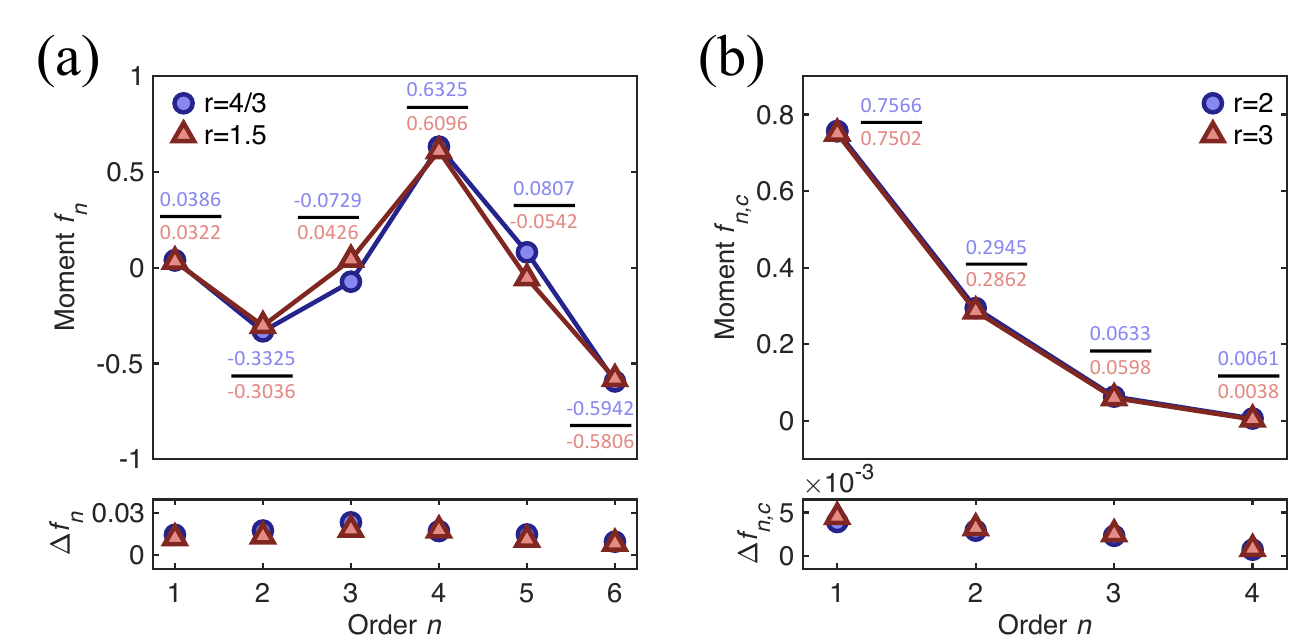}
    \caption{The error-mitigated moments at different noise amplifying factors.
	(a) the results for $2\times2$ lattice of the XY model.(b) the results for 1D 10-qubit closed chain of the TFIM. The standard errors ($\Delta(\cdot)$) are suppressed to around $10^{-2}$ and thus unobservable in this figure.
    }\label{fig:ContrastLZNE}
\end{figure}

To examine the stability of the subset-based noise amplifying method, we take error-mitigated moments of the 1D 10-qubit closed chain of TFIM at $r=1,2$ and $r=1,3$, and $2\times2$ lattice of XY model at $r=1,4/3$ and $r=1,1.5$, respectively (shown in Figure~\ref{fig:ContrastLZNE}). The approximate agreement of the two sets of results obtained with different amplifying factors supports the good linearity of the experiment results relative to the noise amplifying factor and the rationality of our error-mitigated results.

\medskip 
\subsection{Expansion cutoff}
A larger expansion cutoff $N$ improve the convergence of the density-of-states approximated by the kernel-corrected Fourier series. Since the contribution of higher-order moments is suppressed by the Jackson kernel especially when they are quite small, even keeping the first few moments only produces accurate estimates for the thermodynamic quantities.

\subsection{Time step in digital simulation}\label{subsubsec:TimeStep}
\begin{itemize}

    \item Choose the number of time steps $M=1$ when measuring moments of density of states (DOS) to ensure the linearity of expectation values with respect to noise amplification factor $r$, along with the negligible
    trotterization error in digital compilation, which is justified by the differences between exact moments and trotter-moments obtained by theoretical simulation.
  
    \item Choose the number of time steps $M=n$ when measuring a Pauli observable in order to reduce the trotterization error since these moments are quite small. 

\end{itemize}

\subsection{Sampling circuit}
\begin{figure}[htp]
    \centering
    \includegraphics[width=0.25\textwidth]{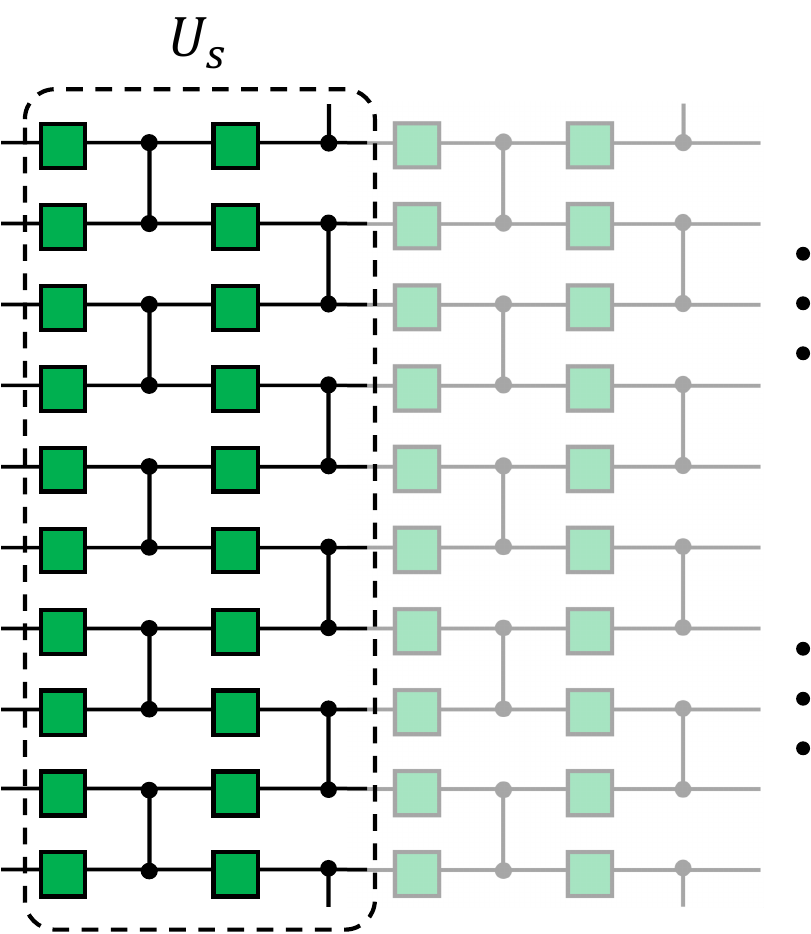}
    \caption{Sampling circuit for measuring a Pauli observable.
	This more intricate circuit architecture is evidently capable of hastening the convergence of the sampling average; at the same time, it concurrently poses a greater susceptibility to experiment noise.
    }\label{fig:DeepSamplingCircuit}
\end{figure}

\begin{itemize}
	\item The sampling average of the DOS moments converges quickly. We only introduce one layer of uniformly sampled single-qubit Clifford gates for the random sampling.
	\item As for the Pauli observable, the sampling circuit should be extended into successive blocks $U_\text{s}$, comprising alternating layers 
	of single-qubit gates and CZ gates, as schematized in panel \textbf{c} of Figure~\ref{fig:DeepSamplingCircuit}.
\end{itemize}

\subsection{Post-selection}
When measuring expansion moments of the $3\times3$ TFIM in digital simulation and the $2\times2$ XY model in hybrid simulation, 
we encounter instability problems in using LZNE.  
Some error-mitigated results $\langle O\rangle_{\text{LZNE}}$ have absolute values larger than $1$, which should be strictly prohibited.   
We use 
{Median Absolute Deviation} (MAD) method to eliminate these outliers~\cite{ZNE1experiment}. 
This means given a time series 
expectation values $\left\{\langle O\rangle_{\text{m}}\right\} = \left\{\langle O\rangle_\text{m,1},\langle O\rangle_\text{m,2},\cdots\right\}$,
we identify the $i$-th one as an outlier when
\begin{equation}
    \left|\langle O \rangle_{\text{m},i} - \text{median}\left(\left\{\langle O\rangle_{\text{m}}\right\}\right)\right| > 2\bar{\sigma},
\end{equation}
where $\bar{\sigma} = k\cdot\text{MAD}$ describe the deviation of the expectation values with 
\begin{equation}
	\text{MAD} = \text{median}\left(\left|\left\{\langle O\rangle_{\text{m}}\right\} - \text{median}\left(\left\{\langle O\rangle_{\text{m}}\right\}\right)\right|\right),
\end{equation}
and $k=1.4826$ assuming normal distribution.

\subsection{Analytical differentiation}
\label{subsubsec:Diff}
The QKFE algorithm allows for the explicit derivation of the analytical form of the free energy and higher order quantities. The two compiling protocols have different expansion forms:
\begin{itemize}
	\item With the virtual-copy compiling protocol, the free energy is expanded as
	\begin{equation}
	F(T) = -\frac{1}{2\beta} \log D^2\left(h_0 f_{0,c}\frac{\sinh(x)}{x} + \sum_{n=1}^N (-1)^n h_n f_{n,c} \frac{ 2x\sinh(x)}{n^2\pi^2 + x^2}\right),\label{eq:VCanaF}
	\end{equation}
	\item With the reference state compiling protocol, the free energy is expanded as
	\begin{equation}
	F(T) = -\frac{1}{\beta} \log D\left(h_0 f_{0}\frac{1-e^{-x}}{x} + \sum_{n=1}^N h_n f_{n} \frac{1-(-1)^n e^{-x}}{\frac{n^2\pi^2}{x} + x}\right),\label{eq:RSanaF}
	\end{equation}
\end{itemize}
where $x \equiv \beta W$, $h_n$ represents the Jackson kernel, $f_{n}$ the expansion moment for an individual system, and $f_{n,c}$ for the doubled free energy.
Higher-order thermodynamic quantities, such as entropy or heat capacity, can be directly obtained by analytically differentiating Eq.~\eqref{eq:VCanaF}~\eqref{eq:RSanaF},
\begin{equation}
S(T) = -\frac{\text{d} F(T)}{\text{d} T}, \quad C(T) = T\frac{\text{d} S(T)}{\text{d} T}.
\end{equation}

\subsection{Technical details of the superconducting quantum processor}\label{ProcessorDetails}
The experiments are performed on three quantum processors: two of them are {\it Zuchongzhi 2.2} , which we will refer to as Chip A and Chip B, and one is {\it Zuchongzhi 3.2}.
Chip A is employed to investigate the one-dimensional 
Transverse Field Ising Model. However, due to suboptimal performance in certain qubits and couplers on Chip 
A, Chip B was subsequently used for quantum simulations of two-dimensional TFIM and XY models. To improve system performance,{\it Zuchongzhi 3.2} is further used  for quantum simulations of two-dimensional TFIM.
Fig.~\ref{T1_readout1} \textbf{(a)} presents the characteristic relaxation 
times of the 
qubits used on three chips, with median values of 30.1 $\mu s$ , 28.7 $\mu s$ and  67.2 $\mu s$, 
respectively.

\begin{figure*}[htbp]
	\begin{center}
		\includegraphics[width=0.8\linewidth]{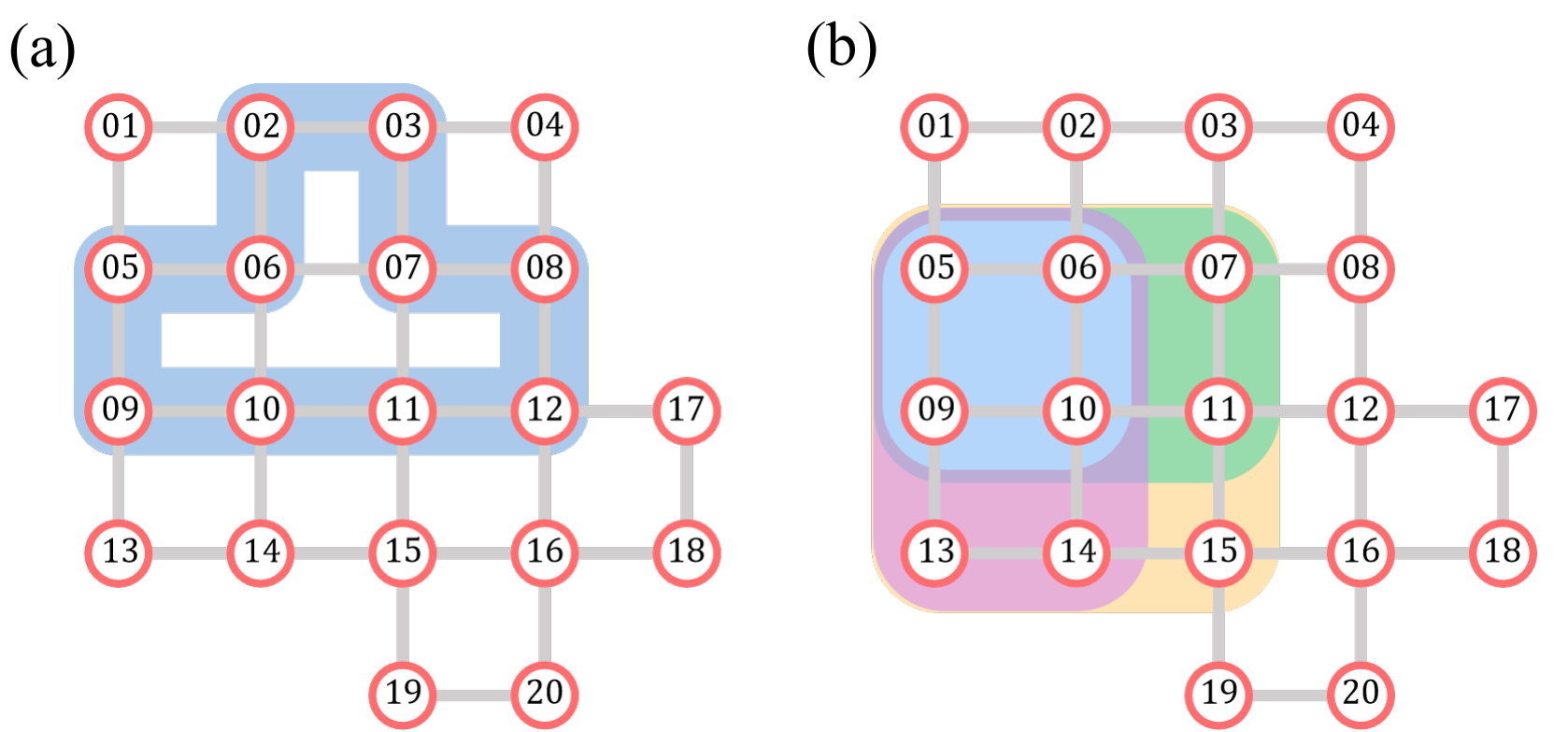}
	\end{center}
	\setlength{\abovecaptionskip}{0pt}
	\caption{\textbf{Chip layout}. \textbf{(a)}  The layout of Chip A. The 
	qubits used for the 1D TFIM model are highlighted  in the blue area. 
	\textbf{(b)}  The  layout of Chip B. The qubits used for the  $3\times3$ 
	and 
		 $2\times2$ 2D TFIM models are highlighted  in the orange area and blue 
		area, respectively. For the 
		 $2\times3$ lattice, purple region  represents the qubits used  for 
		analog simulation while green region  is  designated for digital 
		simulation.}
	\label{processor}
\end{figure*}

The system calibration steps follow the same procedure outlined in our previous works~\cite{zhao2022}. The readout errors are benchmarked, yielding a median 
single-qubit readout error of 
$1.65\%$ on Chip A , $1.85\%$ on Chip B and $1.18\%$ on {\it Zuchongzhi 3.2}. Subsequently, the error rates of the 
single-qubit and two-qubit CZ gates are benchmarked. Fig.~\ref{1q2qxeb1} 
shows  the 
integrated histograms of fidelities of single-qubit Pauli error rates (with 
average errors of $0.102\%$ on Chip A, $0.127\%$ on Chip B and $0.058\%$ on {\it Zuchongzhi 3.2}) and two-qubit 
Pauli error rates(with average errors of $0.479\%$ on Chip A, $0.676\%$ on  Chip B  and $0.377\%$ on {\it Zuchongzhi 3.2}).

Chip B is employed for the two-dimensional quantum simulations of the TFIM and XY models. Because the 
maximum frequency  of  $Q_{11}$ is much lower than that of other qubits, an 
additional 6 qubits are used for the  $2\times3$ lattice analog simulation to 
mitigate the effect of dephasing. The performance of the processor is detailed  
in Table~\ref{tab:4},~\ref{tab:6} and ~\ref{tab:9} for the 
 $2\times2$, $2\times3$ and $3\times3$ lattice configurations, respectively.


\begin{table}[htbp!]
	\centering
    \scalebox{0.8}
    {
	\begin{tabular}{c| c c c c }
	 \hline	
			&$\text{Q}_{05}$ &$\text{Q}_{06}$  & $\text{Q}_{09}$ & 
			$\text{Q}_{10}$\\
			\hline
			$\omega_{\textrm{q}}^{\textrm{max}}/2\pi$~(GHz)
		 	& 5.463    &5.474    &5.363    &5.278\\
		 				\hline	
			$\omega_{\textrm{q}}^{\textrm{idle}}/2\pi $~(GHz)
			& 5.126    &5.306    &5.246    &5.117\\
						\hline	
			$T_1$~($\mu$s)
		    & 39.4  & 39.3&  29.3&  26.7\\
		    			\hline	
			$T_{2e}$~($\mu$s)
			& 3.5&  5.2&    6.2&    5.7\\	
						\hline		
			$f_{00}$~
			&0.992	&0.998	&0.981	&0.992\\
					\hline	
			$f_{11}$~
			 & 0.976 &	0.980	&0.979	&0.982\\
			\hline
			$1$Q XEB fidelity~($\textrm{\%}$)
			&99.86&   99.88&   99.91&   99.90\\
						\hline	
			$1$Q SPB fidelity~($\textrm{\%}$)
			&99.86&   99.88&   99.92&   99.90\\
			\hline			
			&$\text{Q}_{05}$$\text{Q}_{06}$ 
			&$\text{Q}_{05}$$\text{Q}_{09}$ 
			&$\text{Q}_{06}$$\text{Q}_{10}$
			&$\text{Q}_{09}$$\text{Q}_{10}$ \\
			\hline
			$J/2\pi $~(MHz)
		 	&-3.005    &-2.988    &-3.002   &-3.016\\
			\hline
		\end{tabular}
        }
	\caption{
		{\textbf{Parameters of the device in the  $2\times2$ system:} 
		$\omega_{\text{q}}^{\text{max}}/2\pi$ represents the maximum frequency 
		of the 
		qubit; $\omega_{\text{q}}^{\text{idle}}/2\pi$ denotes the idle 
		frequency of 
		the qubit; $T_1$ refer to the energy relaxation time; $T_{2e}$ 
		represents the dephasing time measured via a Hahn echo experiment; 
		$J/2\pi $ represents the coupling strength of the corresponding 
		qubit pair, measured at the working frequency; 
		$f_{11}$ ($f_{00}$) denotes the probability of correctly identifying 
		the 
		qubit state when initially prepared in $|1\rangle$ ($|0\rangle$); 
		The single-qubit cross-entropy benchmarking ($1$Q XEB) fidelity 
		represents the average gate fidelity of the
		$\text{X}/2$ gate measured at idle frequency. The $1$Q SPB fidelity 
		describes the effect of decoherence on the single-qubit $\text{X}/2$ 
		gate. The duration of the $\text{X}/2$ gates for all qubits is $25$~ns. 
		}
	}
	\label{tab:4}	
\end{table}

\begin{table*}[htbp!]

	\centering
    \scalebox{0.8}
    {
		\begin{tabular}{c |c c c c c c c }
			\hline
			\hline	
			&$\text{Q}_{05}$ &$\text{Q}_{06}$    
			&$\text{Q}_{09}$ & $\text{Q}_{10}$   
			&$\text{Q}_{13}$ &$\text{Q}_{14}$   \\
			\hline
			$\omega_{\textrm{q}}^{\textrm{max}}/2\pi$~(GHz)
			& 5.463	&5.474	&5.363	&5.278	&5.368	&5.297	\\
			\hline	
			$\omega_{\textrm{q}}^{\textrm{idle}}/2\pi $~(GHz)
			& 5.143	&5.302	&5.279	&5.104	&5.072	&4.893	\\
			\hline	
			$T_1$~($\mu$s)
			& 27.0	&35.0	&26.0	&41.3	&26.8	&24.7	 \\
			\hline	
			$T_{2e}$~($\mu$s)
			&3.6	&5.9	&7.0	&5.9	&4.6	&3.5 \\
			\hline		
			$f_{00}$
			&0.994	&0.991	&0.972	&0.979	&0.983	&0.996	 \\
			\hline	
			$f_{11}$
			&0.990	&0.985	&0.962	&0.964	&0.963	&0.977	 \\
			\hline
			$1$Q XEB fidelity~($\textrm{\%}$)
			&99.83	&99.83	&99.80	&99.85	&99.88	&99.84	 \\
			\hline	
			$1$Q SPB fidelity~($\textrm{\%}$)
			&99.85	&99.87	&99.85	&99.86	&99.92	&99.86	\\
			\hline			
			&$\text{Q}_{05}$$\text{Q}_{06}$ 
			&$\text{Q}_{09}$$\text{Q}_{10}$
			&$\text{Q}_{13}$$\text{Q}_{14}$
			&$\text{Q}_{05}$$\text{Q}_{09}$ 
			&$\text{Q}_{13}$$\text{Q}_{09}$
			&$\text{Q}_{06}$$\text{Q}_{10}$
			&$\text{Q}_{14}$$\text{Q}_{10}$\\
			\hline
			$J/2\pi $~(MHz)
			&-1.490    &-1.486    &-1.504    &-1.513 &-1.504    &-1.517    
			&-1.506   \\
			\hline
		\end{tabular}
    } 
	\caption{
		{\textbf{Parameters of the device in the  $2\times3$ system.} }
	}
	\label{tab:6}	
\end{table*}

\begin{table*}[htbp!]
	\centering
    \scalebox{0.7}
    {
		\begin{tabular}{c |c c c c  c  c c c c c c c  c}
			\hline
			\hline	
			&$\text{Q}_{05}$ &$\text{Q}_{06}$  & $\text{Q}_{07}$  
			&$\text{Q}_{09}$ & $\text{Q}_{10}$  &$\text{Q}_{11}$  
			&$\text{Q}_{13}$ &$\text{Q}_{14}$  &$\text{Q}_{15}$ \\
			\hline
			$\omega_{\textrm{q}}^{\textrm{max}}/2\pi$~(GHz)
			& 5.463	&5.474	&5.595	&5.363	&5.278	&5.096	&5.368	&5.297 
			&5.391 \\
			\hline	
			$\omega_{\textrm{q}}^{\textrm{idle}}/2\pi $~(GHz)
			& 4.870	&4.964	&4.851	&4.939	&4.810	&4.906	&5.010	&4.888	
			&4.827 \\
			\hline	
			$T_1$~($\mu$s)
			& 35.2	&37.8	&29.6	&30.2	&37.1	&17.3	&39.7	&44.5	
			&27.3 \\
			\hline	
			$T_{2e}$~($\mu$s)
			&2.5	&4.1	&2.3	&4.0	&3.5	&5.3	&3.9	&3.4	
			&3.4 \\
			\hline		
			$f_{00}$
			&0.994	&0.987	&0.975	&0.985	&0.980	&0.998	&0.992	&0.975	
			&0.983 \\
			\hline	
			$f_{11}$
			&0.987	&0.971	&0.968	&0.971	&0.959	&0.965	&0.979	&0.968	
			&0.976 \\
			\hline
			$1$Q XEB fidelity~($\textrm{\%}$)
			&99.84	&99.83	&99.80	&99.83	&99.86	&99.82	&99.87	&99.85	
			&99.83 \\
			\hline	
			$1$Q SPB fidelity~($\textrm{\%}$)
			&99.87	&99.83	&99.83	&99.87	&99.89	&99.85	&99.88	&99.90	
			&99.86 \\
			\hline			
			&$\text{Q}_{05}$$\text{Q}_{06}$ 
			&$\text{Q}_{06}$$\text{Q}_{07}$ 
			&$\text{Q}_{09}$$\text{Q}_{10}$
			&$\text{Q}_{10}$$\text{Q}_{11}$
			&$\text{Q}_{13}$$\text{Q}_{14}$
			&$\text{Q}_{14}$$\text{Q}_{15}$
			&$\text{Q}_{05}$$\text{Q}_{09}$ 
			&$\text{Q}_{13}$$\text{Q}_{09}$
			&$\text{Q}_{06}$$\text{Q}_{10}$
			&$\text{Q}_{14}$$\text{Q}_{10}$
			&$\text{Q}_{11}$$\text{Q}_{07}$
			&$\text{Q}_{15}$$\text{Q}_{11}$\\
			\hline
			$J/2\pi $~(MHz)
			&-1.406    &-1.409    &-1.412    &-1.415 &-1.408    &-1.416   
			&-1.413    &-1.405 &-1.404 &-1.407 &-1.409 &-1.400 \\
		\hline
		\end{tabular}	
    }
	\caption{
		{\textbf{Parameters of the device in the  $3\times3$ system. The 
		frequencies of couplers remain untuned.} }
	}
	\label{tab:9}	
\end{table*}

\begin{figure*}[htbp]
\begin{center}
\includegraphics[width=0.8\linewidth]{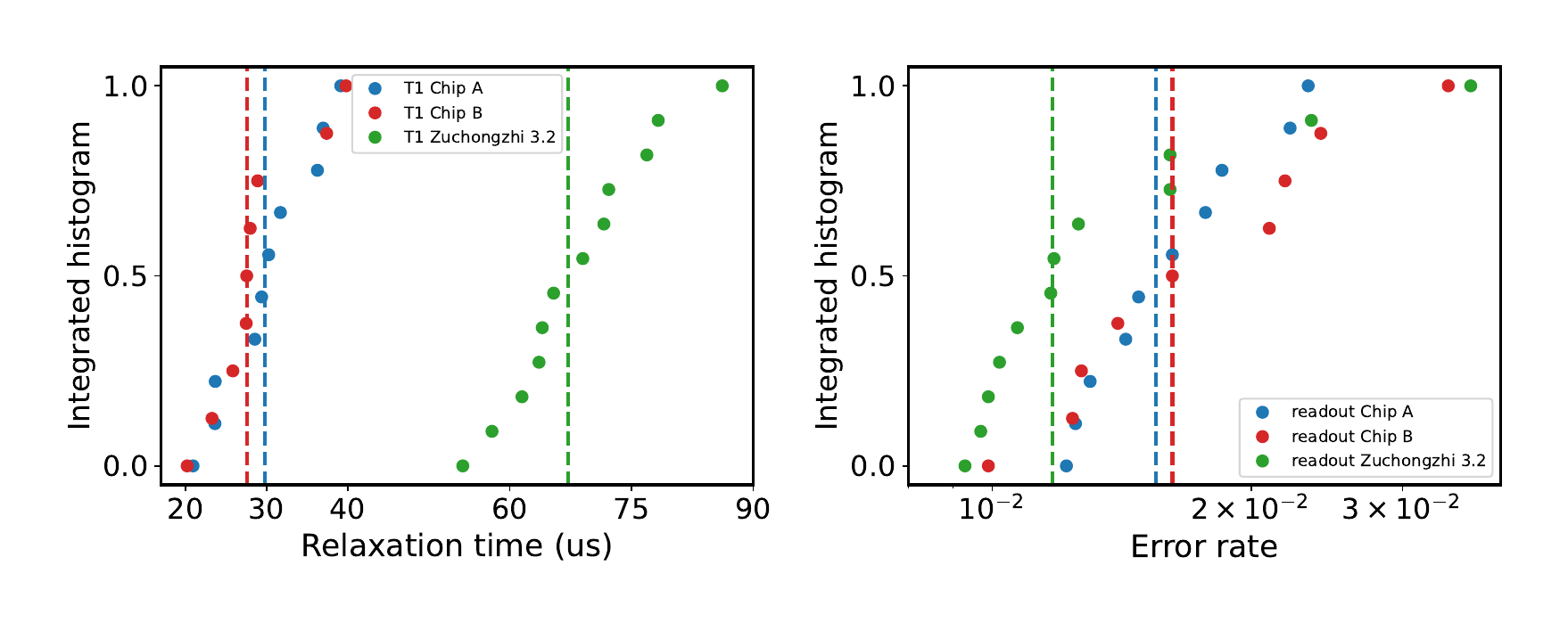}
\end{center}
\setlength{\abovecaptionskip}{0pt}
\caption{\textbf{T1 and readout errors in digital simulation.} \textbf{(a)}  
Integrated histogram of  
T1 of used qubits in two superconducting quantum processors with a dashed line indicating median values of  30.1 $\mu s$ , 28.7 $\mu s$  and 67.2 $\mu s$, respectively. 
\textbf{(b)}  Integrated histogram of the readout error with the median values  of  $1.65\%$ , $1.85\%$ and $1.18\%$.}
\label{T1_readout1}
\end{figure*}

\begin{figure*}[htbp]
	\begin{center}
		\includegraphics[width=0.8\linewidth]{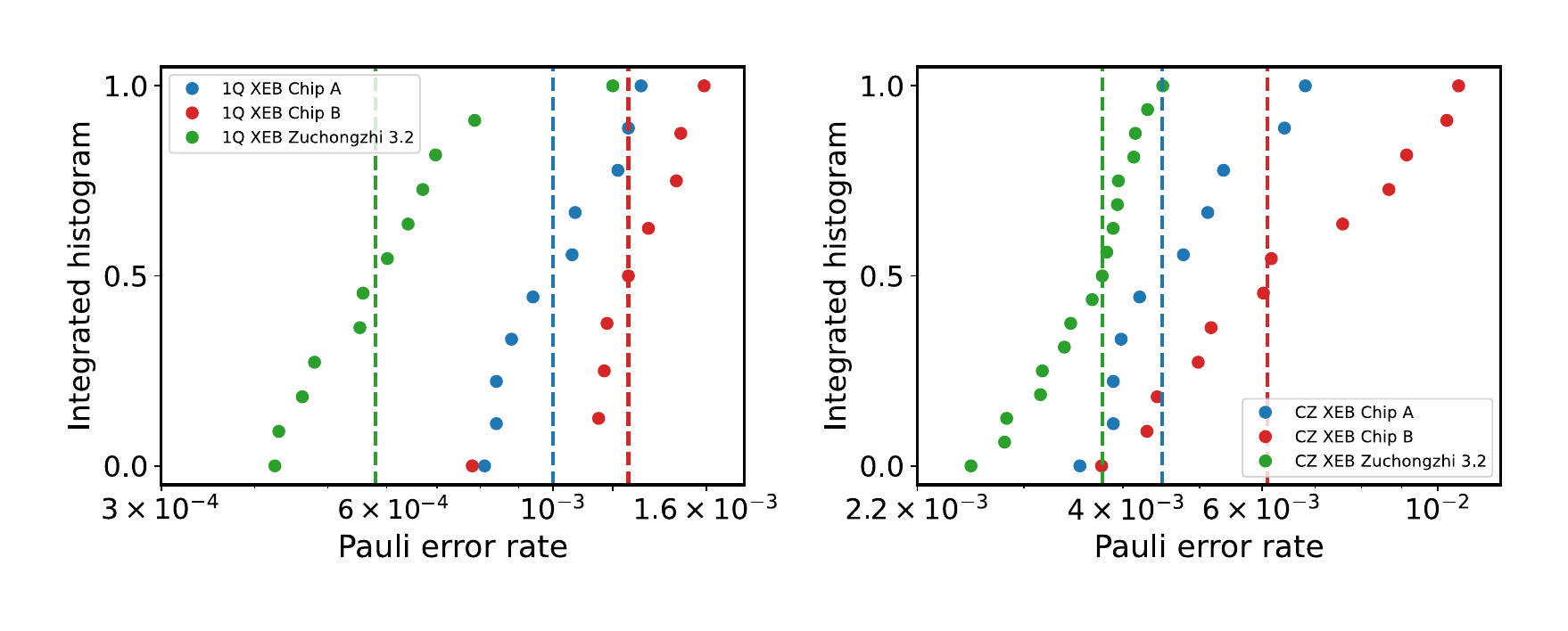}
	\end{center}
	\caption{\textbf{Single-qubit pauli error and CZ pauli error in digital 
	simulation.} \textbf{(a)} 
	Integrated histogram of the single-qubit Pauli error. The median values 
	are $0.102\%$ , $0.127\%$ and $0.058\%$. \textbf{(b)}  Integrated histogram of the 
	two-qubit CZ 
	gate Pauli error.The median values are $0.479\%$ , $0.676\%$ and $0.377\%$.}
	\label{1q2qxeb1}
\end{figure*}

\begin{figure*}[htbp]
	\begin{center}
		\includegraphics[width=0.87\linewidth]{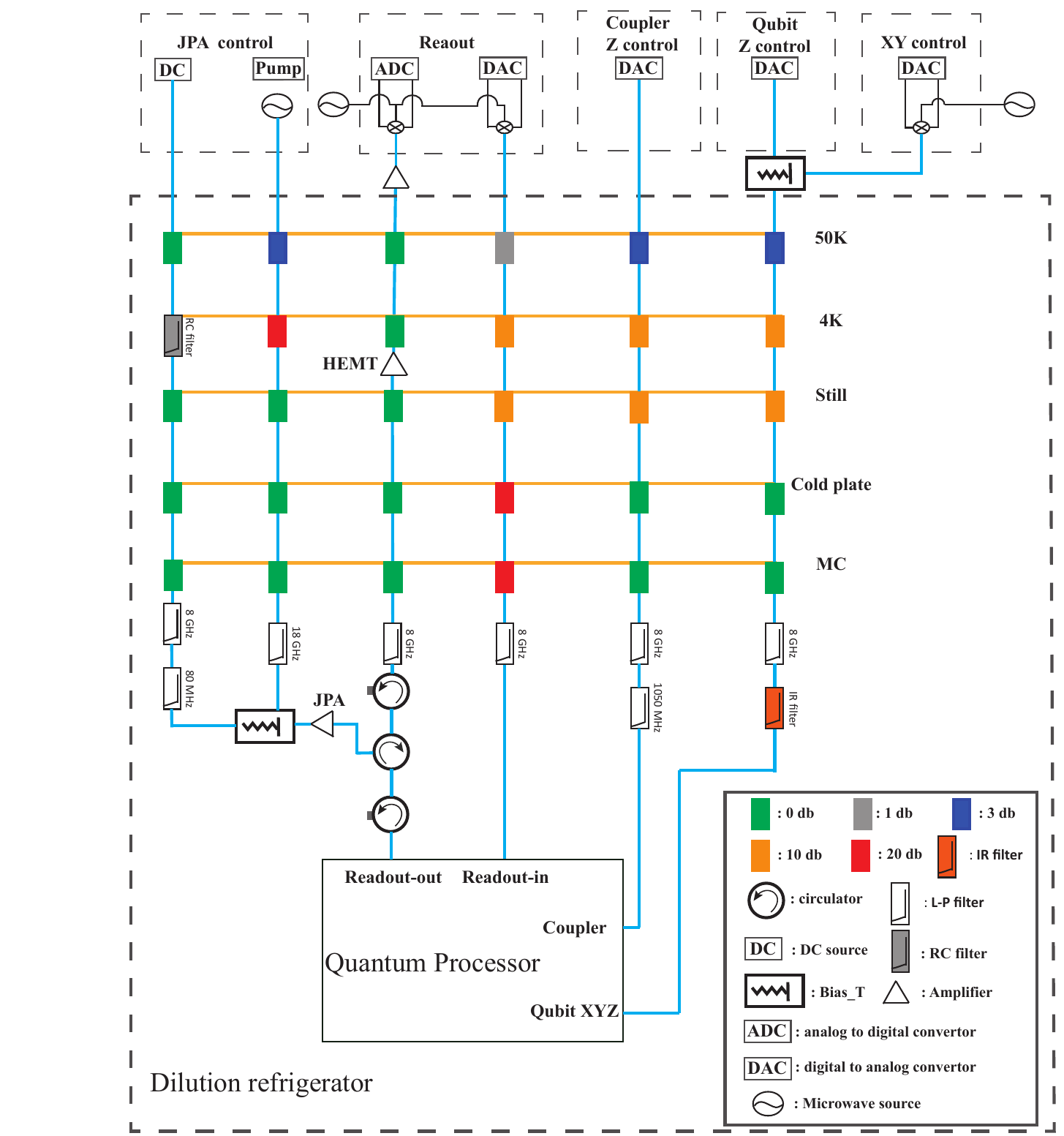}
	\end{center}
	\caption{The schematic diagram of control electronics and wiring. } 
	\label{wiresetup}
\end{figure*}

As illustrated in Fig.~\ref{wiresetup}, our measurement system  of  {\it Zuchongzhi 2.2} comprises a 
dilution refrigerator, control electronics, and wiring. The quantum processor 
is mounted at the base temperature stage of the dilution refrigerator to ensure 
optimal operating conditions. For the qubit XYZ control line, a 23-dB 
attenuator and an infrared filter are installed. The qubit Z control line and 
XY control line are combined together via a bias tee at room temperature. A 
61-dB attenuator is installed for the readout-in line, while a 23-dB attenuator 
is employed for the coupler Z control line and JPA pump line. Additionally, to 
mitigate high-frequency noise, low-pass filters are installed at the mixing 
chamber stage for various lines, including the qubit XYZ control line, 
readout-in line, readout-out line, coupler Z control line, JPA pump line, and 
JPA DC line. Specifically, for the JPA dc control line, a RC filter with a 10 
kHz cutoff frequency is installed at the 4 K stage. Finally, the readout 
signal is initially amplified by a high electron mobility transistor (HEMT) 
amplifier situated at the 4 K stage, followed by further amplification using a 
room temperature amplifier.

\subsection{Calibration details in analog simulations}
\label{sec:Calibration}
Here, we provide detailed calibration procedures for our analog 
simulation experiments.

\begin{figure*}[htbp]
	\includegraphics[width=0.85\linewidth]{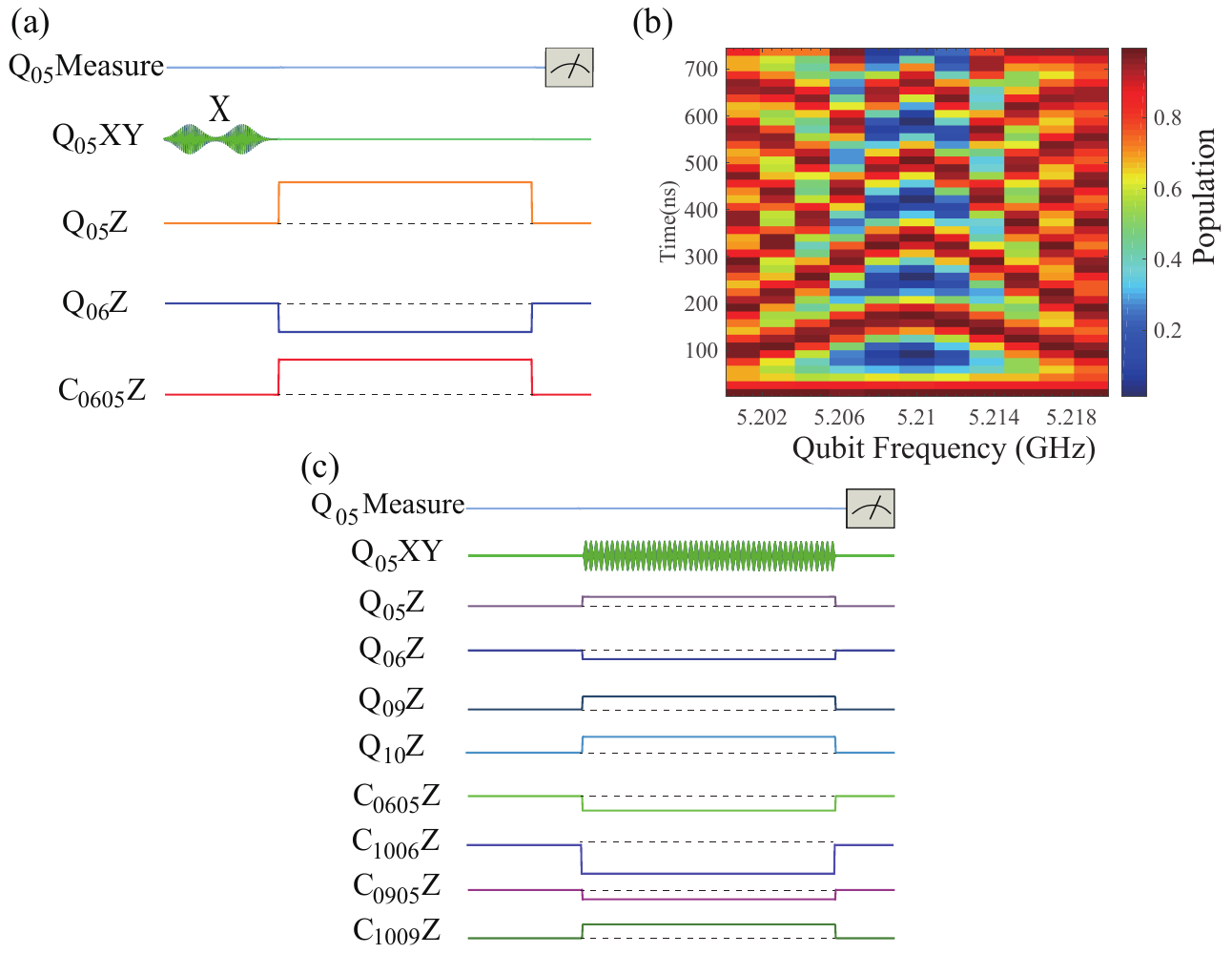}
	\caption{\textbf{(a)} The sequences of swap experiment for $Q_{05}$ and 
	$Q_{06}$, while 
	other qubits are tuned ~150~MHz away from working frequency. 
	\textbf{(b)} Typical 
	oscillation between two qubits $Q_{10}$ and $Q_{06}$ in the  $2\times2$
	lattice. The coupling strength is about -3.0~MHz. \textbf{(c)} The 
	sequences of 
	spectroscopy experiment. } 
	\label{fig:figs2_swap} 
\end{figure*}

\emph{Calibration of coupling strength and working frequency.---}
In our work, we align all qubits to the same working frequency and adjust all 
couplers to achieve the required  coupling strengths. To ensure precise coupling strengths and consistent working 
frequencies, we first 
calibrate the $Z$ pulse distortion, followed by 
testing the $Z$ pulse crosstalk. The maximum absolute 
value 
of $Z$ pulse crosstalk between qubits is below 0.17$\%$ and we calibrate this 
crosstalk accordingly.  
Although $Z$ pulse crosstalk from qubits to couplers can influence the 
coupler frequencies and thereby  affect coupling strength, this effect is 
second-order and thus not calibrated. Similarly, $Z$ pulse crosstalk between 
couplers is not calibrated.

Since coupling strengths are related to qubit frequencies, we measure the 
coupling strength when  detuning the qubits to their working frequencies during 
the swap experiment,as shown in 
Fig.~\ref{fig:figs2_swap} \textbf{(a)}. For the  $2\times2$ ,  $2\times3$ 
lattice and  $3\times3$ lattice, the working frequencies are
5.207 $\text{GHz}$, 5.223 $\text{GHz}$ and 4.873 $\text{GHz}$, respectively.

The $Z$ pulse applied to couplers can affect  qubit frequencies in two primary  
ways:
\begin{itemize}
	\item[(1)] The interaction between a specific coupler and one of its 
	nearest qubits results in a change in qubit frequency when the coupler’s 
	frequency varies, representing a nonlinear effect.
	\item[(2)]  Linear crosstalk from couplers' control lines to qubits' 
	control lines.
\end{itemize}
To mitigate both nonlinear effects and linear crosstalk, all couplers are 
detuned to their calibrated frequencies during our spectroscopy experiment. 
This approach ensures accurate calibration of qubit working frequencies, while 
other qubits are detuned away from the working frequency.The calibration 
sequences are illustrated in Fig.~\ref{fig:figs2_swap} \textbf{(c)}. 

For the $3\times3$ lattice, the error mitigation of noise is 
achieved through the utilization of $Z$ and $I$ gates, as described in 
Ref.~\cite{MITZgate}. Given the intricate implementation of $Z$ gates,the 
frequencies of couplers remain unaltered throughout the entire analog 
simulation.

\begin{figure*}[htbp]
		\begin{center}
	\includegraphics[width=0.85\linewidth]{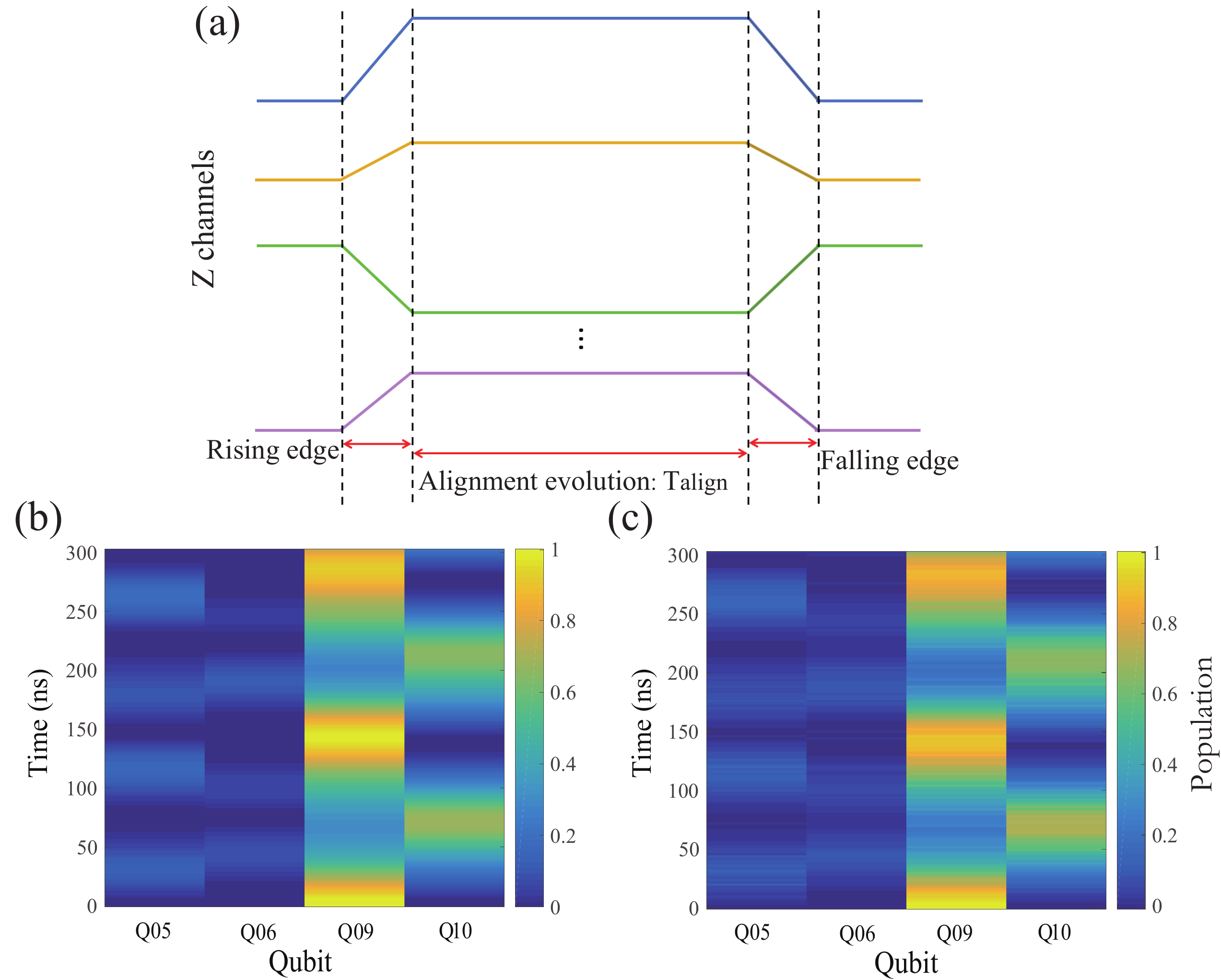}
		\end{center}
	\caption{\textbf{(a)} Experimental evolution time of analog
	simulation.
	\textbf{(b)} The numerical simulation results of the time evolution of 
	$\langle \hat{n}_{i}\rangle$ when $Q_{09}$ is excited in the  $2\times2$ 
	lattice. 
	\textbf{(c)} The experiment results of the data in \textbf{(b)}.} 
	\label{quantum_walks} 
\end{figure*}

\emph{Fine calibration of working frequency and effective evolution time.---}
As shown in Fig.~\ref{quantum_walks} \textbf{(a)}, our evolution encompasses the 
rising edge, falling edge and alignment evolution. Residual coupling during the 
rising
and falling edges may affect the evolution. We utilize  
multi-qubit excitation propagation to calibrate both the frequency drift and 
the effective evolution time. The calibration process is listed below:
\begin{itemize}
	\item[(1)] Initialize $Q_m$ among the $N$ qubits to $|1\rangle$ and leave 
	the others in $|0\rangle$. Subsequently, all qubits are tuned to the target 
	frequency $\omega_m$. After an evolution time $T_{align}$, the population of all 
	sites is measured, 
	i.e., $\langle \hat{n}_{i}\rangle = \langle \hat{\sigma}_{i}^{+} 
	\hat{\sigma}_{i}^{-} \rangle$. Here, we set $\omega_m$ as
	\begin{eqnarray}
		\omega_m(m)= \omega_c+s\times k(m),
		\label{alignmentfreq}
	\end{eqnarray}
	where $\omega_c$ is the chosen working frequency, and $s$ represents  
	frequency difference, set to $\pm 4$~MHz in two separate  measurements for 
	the  $2\times2$ lattice. 
	$\textit{k(m)}$ represents the ratio of frequecny difference for specific 
	qubit. For the  $2\times2$ lattice, $k(05)=-2, k(06)=-1, k(09)=1, k(10)=2$.
	Different sites are excited sequentially, preparing 4 initial states for 
	the evolution, which yields 8 time-dependent population distributions, 
	$Z_{\textrm{exp}}$.
	\item[(2)] We use QuTiP~\cite{QuTiP1} to simulate the evolution. The 
	Hamiltonian employed  in the simulation is
	
	\begin{eqnarray}\nonumber
		\hat{H}&=&\hat{H}_{\text{C}}\\\nonumber
		&+&\sum_{n=5,6} \frac{k(n)\times 
		s+\delta_\omega(n)}{2}(1-2\hat{\sigma}_{n}^{-}\hat{\sigma}_{n}^{+})\\\nonumber
		&+&\sum_{n=9,10} \frac{k(n)\times 
		s+\delta_\omega(n)}{2}(1-2\hat{\sigma}_{n}^{-}\hat{\sigma}_{n}^{+})
		\label{H-array}
	\end{eqnarray}
	where $\hat{\sigma}_{n}^{-}$($\hat{\sigma}_{n}^{+}$ ) denotes the 
	annihilation (creation) operator of the $n$-th qubit, $\hat{H}_{\text{C}}$ 
	is the 
	Hamiltonian (\ref{HC}) and $\delta_\omega$ refers to the frequency drifts.
	
		\begin{eqnarray}\nonumber
		\hat{H}_{\text{C}}&=& \sum_{n=5,9} 
		J_{n,n+1}(\hat{\sigma}_{n}^{+}\hat{\sigma}_{n+1}^{-} + \text{h.c.})\\
		&+&\sum_{n=5,6} 
		J_{n,n+4}(\hat{\sigma}_{n}^{+}\hat{\sigma}_{n+4}^{-} + \text{h.c.})
		\label{HC}
	\end{eqnarray}
	For the 8 time-dependent distributions $Z_{\textrm{exp}}(\textrm{ii})$ , 
	the population propagations are numerically simulated using the
	corresponding initial states and $\omega_m$ for various start time. Here, the 
	start time indicates the effective evolution time when
	$T_{align}=0$, from which the expected values 
	$Z_{\textrm{sim}}$ for all qubits are derived. The deviation between the 
	numerical and experimental results is defined as 
	$Z_{\textrm{diff}}(\textrm{ii})=(Z_{\textrm{sim}}-Z_{\textrm{exp}})^2 $, 
	with the overall distance of all $8$ evolutions given by 
	$Z_{\textrm{diffall}}=\sum_{\textrm{ii}=1}^{8} 
	Z_{\textrm{diff}}(\textrm{ii})$. For different start times, we employ the 
	Nelder-Mead optimization algorithm to minimize 
	$Z_{\textrm{diffall}}$, ultimately determining the $\delta_\omega$ array and the 
	optimal start time.
	\item[(3)] Once $\delta_\omega$ is obtained, it is added to $\omega_m$ as an 
	offset calibration to correct the drift. Steps (1)-(3) are then repeated 
	until the absolute values of frequency drift are small.
\end{itemize}

Fig.~\ref{quantum_walks}\textbf{(c)} presents a portion of the calibration 
results for the
 $2\times2$ lattice , where $Q_{09}$ is excited and evolved for about 
$300$~ns. The simulation result in Fig.~\ref{quantum_walks}\textbf{(b)} closely matches the 
experiment result.
The final distance and the sum of the absolute values of  frequency 
drift for different simulation start times are summarized in 
Table~\ref{quantumwalks}. 
Based on these results, we select 0 $\textrm{ns}$ as the optimal start time.

\begin{table}[]
	\centering
	\begin{tabular}{c |c c c c c}
		\hline
		\hline
		Start time ~(ns)
		&$\text{0}$ &$\text{1}$ &$\text{2}$ &$\text{3}$ &$\text{4}$ 	\\
		\hline
		Sum of absolute values of\\
		 frequency drift $/2\pi$~(MHz)
		&0.07	&0.07 &0.07	&0.07	&0.07\\
		\hline
		$Z_{\textrm{diffall}}$ after optimize
		&5.69 &5.99	& 6.85 &8.31  &10.37	\\
		\hline
		Original $Z_{\textrm{diffall}}$ 
		&5.74 &6.03	& 6.90 &8.37  &10.44	\\
		\hline
	\end{tabular}	
	
	\caption{
		{\textbf{Frequency drifts and distance for different simulation start 
		times in the $2\times2$ lattice.}  }
	}
	\label{quantumwalks}	
\end{table}

\begin{figure*}[htbp]
	\includegraphics[width=1.0\linewidth]{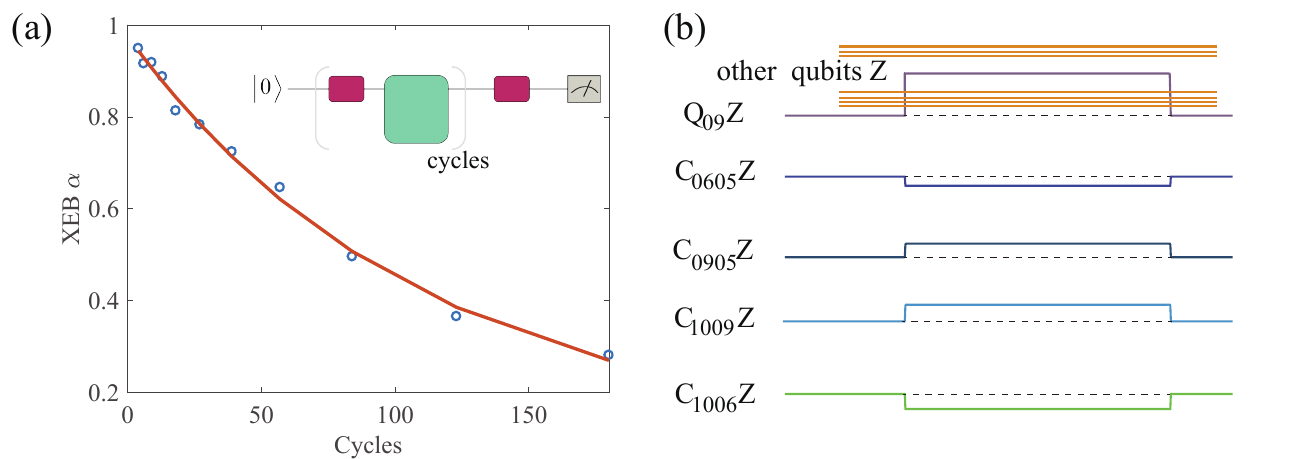}
	\caption{\textbf{(a)} Single-qubit $Z$ gate fidelity versus cycles number 
	for $Q_{09}$ at the second-order moments in the  $2\times2$ 
		lattice. The fidelity of single-qubit $Z$ gate is 
		99.32$\textrm{\%}$. Inset: The red box represents single-qubit gate and 
		the green 
		box represents real $Z$  gate which is described in \textbf{(b)} with 
		more 
		details. \textbf{(b)} Single qubit real $Z$ gate for $Q_{09}$ in the 
		 $2\times2$ lattice. } 
	\label{fig:figs3_rZ} 
\end{figure*}

\emph{Calibration of single qubit $\text{Z}$ gate.---}
In our analog simulation, inverse single-qubit gates must be applied 
before measurement to accelerate the convergence of moments. In the 
experiment, all qubits are detuned from their idle frequencies to the working 
frequency to realize the evolution, which is achieved by applying rectangular 
pulses to each qubit and coupler. This operation results in the accumulation of 
a dynamical phase, which must be canceled by applying rotation pulses after the 
rectangular pulses. To benchmark and calibrate this dynamical phase, we employ 
single-qubit $Z$-gate cross-entropy benchmarking, as shown in 
Fig.~\ref{fig:figs3_rZ}\textbf{(b)}. To isolate the dynamical phase from 
coupling effects and ensure accurate in-situ benchmarking, we measure  the 
dynamical phase  for each qubit individually,when activating all couplers and 
detuning the other qubits away from the working frequency. For the $3\times3$ 
lattice, our test waveform incorporates $Z$ and $I$ gates.

\subsection{Self duality of 1D TFIM}
To highlight the efficiency of the QKFE algorithm, we conducted an additional examination of the self-duality property of the 1D TFIM. Self-duality is not merely a feature of the scaling dimensions or critical exponents; it is an intrinsic algebraic property of specific critical theories~\cite{SelfDuality2}. For the 1D TFIM, which is characterized by the Hamiltonian
\begin{equation}
H(J,g) = -J\sum_j Z_j Z_{j+1} - g\sum_j X_j,
\end{equation}
the self-duality transformation is defined as
\begin{equation}
{\cal D}X_j {\cal D}^{-1} = Z_j Z_{j+1},\quad {\cal D}Z_j {\cal D}^{-1} = \prod_{k\leq j}X_k,
\label{eq:54}
\end{equation}
indicating that the Hamiltonian undergoes a precise transformation, ${\cal D}H(J,g){\cal D}^{-1} = H(g,J)$, for an infinitely long chain. However, the second equality in Eq.~\eqref{eq:54} is problematic at the boundaries, leading to an approximate self-duality transformation for finite-sized models. As a result, this discrepancy yields two approximate equalities concerning the free energy and correlation functions:
\begin{align}
\Tr\left[e^{-\beta H(J,g)}\right] &\simeq \Tr\left[e^{-\beta H(g,J)}\right], \\
\frac{1}{Z[\beta]}\Tr\left[Z_1Z_2e^{-\beta H(J,g)}\right] &\simeq \frac{1}{Z[\beta]}\Tr\left[X_1e^{-\beta H(g,J)}\right].
\end{align}
These equalities can be experimentally verified to substantiate the self-duality of the 1D TFIM\@.
\par
The experimental verification of the self-duality of free energy has been presented in Figure $\bm{3}$ in the main text. 
Figure~\ref{fig:TFIM_Observable} shows the experiment result of the observables in 1D TFIM at the quantum critical point ($g=J$).

\bibliography{references}